\documentclass[aps,prc,superscriptaddress,twoside,twocolumn,nofootinbib,showpacs]{revtex4-1}
\usepackage{amsmath,amssymb}
\usepackage{mathtools}

\usepackage[usenames,dvipsnames]{xcolor}
\usepackage{braket}
\usepackage{graphicx,bm}
\usepackage{slashed}

\usepackage{booktabs}
\usepackage{tensor}
\usepackage{color}
\usepackage{changes}

\usepackage{url}

\allowdisplaybreaks

\begin{document}


\title{\boldmath Nuclear isospin asymmetry in $\alpha$ decays of heavy nuclei}

\author{Eunkyoung \surname{Shin}}
\email{shinek@knu.ac.kr}
\affiliation{Department of Physics, Kyungpook National University, Daegu 41566, Korea}

\author{Yeunhwan \surname{Lim}}
\email{ylim9057@ibs.re.kr}
\affiliation{Rare Isotope Science Project, Institute for Basic Science, Daejeon 34047, Korea}

\author{Chang Ho \surname{Hyun}}
\email{hch@daegu.ac.kr}
\affiliation{Department of Physics Education, Daegu University, Gyeongsan, Gyeongbuk 38453, Korea}

\author{Yongseok \surname{Oh}}
\email{yohphy@knu.ac.kr}
\affiliation{Department of Physics, Kyungpook National University, Daegu 41566, Korea}
\affiliation{Asia Pacific Center for Theoretical Physics, Pohang, Gyeongbuk 37673, Korea}

\begin{abstract}
The effects of nuclear isospin asymmetry on $\alpha$ decay lifetimes of heavy nuclei are
investigated within various phenomenological models of nuclear potential for the $\alpha$ particle.
We consider the widely used simple square well potential and Woods-Saxon potential, and modify them 
by including an isospin asymmetry term. 
We then suggest a model for the potential of the $\alpha$ particle motivated by a microscopic 
phenomenological approach of the Skyrme force model, which naturally introduce the isospin 
dependent form of the nuclear potential for the $\alpha$ particle.
The empirical $\alpha$ decay lifetime formula of Viola and Seaborg is also modified to include 
isospin asymmetry effects.
The obtained $\alpha$ decay half-lives are in good agreement with the experimental data and 
we find that including the nuclear isospin effects somehow improves the theoretical results for 
$\alpha$ decay half-lives. 
The implications of these results are discussed and the predictions on the $\alpha$ decay lifetimes 
of superheavy elements are also presented.
\end{abstract}


\date{\today}

\maketitle




\section{Introduction}

The nuclear $\alpha$ decay has been one of the most important tools to study nuclear forces 
and nuclear structure~\cite{Mang64}. 
Even today, its role cannot be overemphasized in the investigation of nuclear properties and, 
in particular, in identifying syntheses of new elements. (See, for example, Refs.~\cite{GSI,FRIB}.)
Although many facets of the nuclear force were uncovered and understood, there still remain a lot of 
questions to be explored. 
One very naive, but quite nontrivial question would be how many nucleons can aggregate in 
the heaviest nucleus.
Since every nucleus is dynamical and the $\alpha$ decay is one of the major decay processes of 
heavy nuclei, the investigation of $\alpha$ decays of superheavy elements is required to find a clue
to answer this question.

The structure of superheavy elements and their syntheses have been exciting research
topics in both experimental and theoretical nuclear physics~\cite{HM00}. 
These topics attract recent research interests thanks to the construction of new facilities
of rare isotope beams, which will allow the investigation of very neutron-rich nuclei as well as 
superheavy elements.
The stability of nuclei can be achieved through the balance between the attractive nuclear force and 
the repulsive Coulomb force. 
As the number of protons increases, the Coulomb repulsion increases, thus more neutrons are required 
to form a bound state. 
However, the energy of neutron-rich nuclear matter is higher than that of symmetric nuclear matter
because of the nuclear symmetry energy contribution to the total energy. 
Therefore, the nuclear symmetry energy is important to understand the structure of heavy, in particular, 
very neutron-rich nuclei~\cite{LCK08}. 
Furthermore, unstable heavy nuclei eventually decay through spontaneous fission, beta decay, 
nucleon, and $\alpha$ emissions so it deserves to study the role of nuclear symmetry energy or 
the change of nuclear potential due to nuclear isospin asymmetry in these decay processes.

In the standard approach, the $\alpha$ decay lifetimes are governed by the effective potential for the
nuclear force which combines the core nucleus and the $\alpha$ cluster.
There are several phenomenological potential models for explaining the measured data of $\alpha$ 
decay lifetimes, which include the simple $\alpha$ cluster model 
with a square well potential model~\cite{BMP91, BMP92}, $\cosh$-type potential model~\cite{BMP92b}, 
generalized liquid droplet model (GLDM)~\cite{Royer00, DZWZL10}, 
and density-dependent M3Y (DDM3Y) effective interaction~\cite{RSB05, SCB07,RSB08}. 
In the simple cluster model, the $\alpha$ particle is trapped by the core nucleus 
in a nuclear plus Coulomb potential and the $\alpha$ decay 
happens as the bound $\alpha$ particle escapes from the potential barrier by quantum tunneling. 
The shape of the effective nuclear potential felt by the $\alpha$ particle is determined
by fitting the parameters of the potential to the measured $\alpha$ decay lifetimes.
Despite its simplicity, these models are quite successful to describe $\alpha$ decay lifetimes even 
quantitatively~\cite{BMP91,BMP92}.
For a more complete description of the data, one, of course, needs to develop more realistic potential 
models for the $\alpha$ particle.

Improvement of simple potential models has been pursued in several ways.
For example, in the simple potential models illustrated above, the shape of a nucleus is robust and 
does not change during the decay process. 
Therefore, more realistic treatment on the shape evolution was anticipated and investigated, e.g., in the 
GLDM in Refs.~\cite{Royer00, DZWZL10}.
On the other hand, it is also desirable to understand the $\alpha$ potential in nuclear matter from 
a microscopic approach.
Along this direction, the authors of Refs.~\cite{RSB05, SCB07, RSB08} parameterized the $\alpha$ 
particle potential using three Yukawa-type finite range forces that are modified by nuclear density. 
In this approach, it is assumed that the core nucleus follows the Fermi density profile and the 
$\alpha$ particle has the Gaussian density profile.

In the present work, we explore the nuclear isospin asymmetry effects in $\alpha$ decay half-lives 
of heavy nuclei. 
The $\alpha$ potential depth depends on isospin asymmetry~\cite{DI05, LZZS10} and the potential 
depth from isospin asymmetry effects is naturally embedded when double folding model
is employed~\cite{RSB05, SCB07, RSB08}.
The effects of the nuclear symmetry energy, i.e., the nuclear isospin asymmetry effects, 
were also considered in the computation of the $Q$ value of the emergent $\alpha$ particle in 
Refs.~\cite{DZS11,DZG12}.
But the nuclear symmetry energy can affect the $\alpha$ decay lifetimes also through the nuclear 
potential of the $\alpha$ particle.
Therefore, it would be legitimate to investigate the effects of the nuclear symmetry energy on 
$\alpha$ decay lifetimes through the modifications of the effective potential of the $\alpha$ particle,
which may affect, in particular, the $\alpha$ decay half-lives of neutron-rich nuclei.
We will address this issue in the present work.

Recent progress shows that the nuclear symmetry energy is well constrained both by experimental
data and theoretical calculations at least near the normal nuclear density, and its effects have been 
investigated widely in various physical quantities of systems from nuclei to neutron 
stars~\cite{LCK08,RHL11,GCR11,TSCD12, LL12,DSS13}. 
Since far neutron-rich nuclei play a crucial role in understanding of exotic nuclear structure,
it is important to see how the nuclear symmetry energy affects the $\alpha$ decay lifetimes~\cite{LCK08}.
In principle, therefore, the nuclear symmetry energy should be considered in developing the effective
$\alpha$ potential. 
Instead of invoking a complex microscopic calculation, however, we revisit the simple cluster model 
and modify the $\alpha$ particle potential by including the isospin asymmetry term.
The model parameters are then fitted to the existing experimental data and they are used to predict
the lifetimes of unknown elements. 
For a model based on more microscopic approach we also suggest a potential as a functional of proton 
and neutron densities relying on the Skyrme force model. 
In this approach, the isospin asymmetry effects affect both in nuclear potential and proton distribution so
the penetration length depends on the unequal number of neutrons and protons. 
Compared with Yukawa type finite-range double folding model~\cite{RSB05, SCB07, RSB08}, our 
approach is based on zero-range nuclear force (see Appendix).
Finally, we will discuss the modification of the empirical formula of Viola and Seaborg for $\alpha$ decay 
lifetimes by explicitly including the isospin asymmetry term.

This paper is organized as follows. 
In Sec.~\ref{sec:theory}, we discuss the general features of the potential model for nuclear 
$\alpha$ decay.
To investigate the nuclear isospin asymmetry effects, we first consider the square well potential and 
the Woods-Saxon potential, and then suggest a potential motivated by the Skyrme force model
before we discuss the modification of the empirical Viola-Seaborg formula.
The computed $\alpha$ decay lifetimes of heavy nuclei of $Z = 106$--$118$ are compared with 
existing experimental data in Sec.~\ref{sec:results}. 
We also present the predictions of $\alpha$ decay half-lives for superheavy elements in the
range of $Z = 117$--$122$.
Section~\ref{sec:summary} contains the summary and discussion.


\section{\boldmath  Models of $\alpha$  Decay}
\label{sec:theory}

In this section, we briefly review and introduce potential models for the $\alpha$ particle.
The fitting process to find the values of the potential parameters is also shortly described.

\subsection{General features}
\label{subsec:general}

In the present work, with a given model potential, we make use of the Wentzel-Kramers-Brillouin (WKB) 
approximation to calculate the $\alpha$ decay half-lives of heavy nuclei.
In the $\alpha$ cluster model, the $\alpha$ particle interacts with the core nucleus, which becomes
the daughter nucleus after decay, through the strong nuclear interaction and Coulomb interaction. 
Even though the $\alpha$ particle has a finite size ($r_{\alpha}^{} \sim 2.0 $ fm), it is negligibly small 
since its volume fraction to the decaying nucleus, or mother nucleus, is less than $1/25$ if $A \ge 100$. 
This justifies the approximation of treating the $\alpha$ particle as an elementary particle, 
and the $\alpha$ decay can be described by the quantum tunneling of a pointlike particle.
The potential of the $\alpha$ particle produced by the core nucleus can be written as
\begin{equation}
V_\alpha (r) = V_N(r) + V_C(r) + V_L(r)\, ,  
\label{eq:pot}
\end{equation}
where $V_N(r)$ is the nuclear potential, $V_C(r)$ is the Coulomb potential, and $V_L(r)$ is the 
centrifugal barrier.
The explicit forms of each potential are model-dependent and will be discussed later in this section.

In the semi-classical approximation~\cite{BMP91}, the half-life is computed as
\begin{equation}
T_{1/2} = \frac{\hbar \ln2}{\Gamma}
\end{equation}
with~\cite{GK87b}
\begin{equation}
\Gamma = \mathcal{P} \mathcal{F} \frac{\hbar^2}{4m_\mu}
 \exp\left[-2\int_{r_2^{}}^{r_3^{}} dr\, k(r) \right] ,
\end{equation}
where $m_\mu^{}$ is the reduced mass of the system and $\mathcal{P}$ is the $\alpha$ 
particle preformation probability.
It is the probability that an $\alpha$ particle is formed inside a nucleus, so that the $\alpha$ decay
is described as the emission of the preformed $\alpha$ particle.
It is understood to be similar to the spectroscopic factor of protons in the case of proton emission 
process~\cite{ASN97}.
Since our purpose is to see the effects of nuclear isospin asymmetry in the nuclear potential of
the $\alpha$ particle and the parameters of the potential will be fitted by experimental data, 
we simply assume $\mathcal{P} = 1$ throughout this work.
The normalization factor $\mathcal{F}$ is defined by
\begin{equation}
\mathcal{F} \int_{r_1^{}}^{r_2^{}} \frac{dr}{k(r)} \cos^2 
\left[ \int_{r_1^{}}^{r_2^{}} dr^\prime \,k(r^\prime) - \frac{\pi}{4} \right] = 1\,.
\end{equation}
Physically, $\mathcal{F}$ is the assaulting frequency of the $\alpha$ particle to the potential well 
by the core nucleus. 
Here, $r_1^{}$, $r_2^{}$, and $r_3^{}$ denote classical turning points at the centrifugal barrier, inner 
and outer barriers of the Coulomb potential, respectively.
The wave number of the $\alpha$ particle is given by
\begin{equation}
k(r) = \sqrt{\frac{2m_\mu}{\hbar^2}\left\lvert Q_\alpha - V(r) \right\rvert}\, ,
\end{equation}
where  $Q_\alpha$ is the energy of the system during the decay process. 
It is known that the $\alpha$ decay half-lives are very sensitive to the value of $Q_\alpha$.

To compute the $\alpha$ decay half-life, one has to model the potential appearing in Eq.~(\ref{eq:pot}).
The potentials to model the interaction between the $\alpha$ particle and the core nucleus are 
parameterized and these parameters are determined by the Monte Carlo method with which we 
minimize the root-mean-square (rms) deviation $\sigma$ defined as
\begin{equation}
\sigma = \sqrt{\frac{1}{N_{\rm data}-1} \sum 
\left(\log_{10} \frac{T_{1/2}^{\text{theor}}}{T_{1/2}^{\text{exp}}}\right)^2}\,,
\label{eq:rms}
\end{equation}
where $N_{\rm data}$ is the total number of data.
In the present work, we consider three models for the potential 
and explore a possible modification
of the empirical Viola-Seaborg formula for $\alpha$ decay half-lives.

\subsection{\boldmath Square well potential for $\alpha$ particle}

We first consider the square well potential as the simplest choice for the nuclear potential 
of the $\alpha$ particle.
In Ref.~\cite{BMP91}, it was found that the square well potential approach is quite successful
to explain the $\alpha$ decay half-lives considering its simplicity.
The square well potential assumes that nuclei have sharp edges as in the liquid drop model. 
Since a uniform density is assumed, the nuclear potential for the $\alpha$ particle is constant and 
attractive.
Therefore, we have 
\begin{equation}
V_N(r) = \left\{ \begin{array}{ll} V_0 & \mbox{ for } r < R, \\
0  & \mbox{ for } r \ge R, 
\end{array} \right.
\end{equation}
where $R$ is the
radius of the core nucleus and $V_0 < 0$.
Since our aim is to explore the effects of the nuclear isospin asymmetry, we just follow the square 
well potential model of Refs.~\cite{BMP91, BMP92}, which assumes the Coulomb potential of the 
surface charge form as
\begin{equation}
V_C(r) = 
\begin{cases} \displaystyle
\frac{ Z_1 Z_2e^2}{R}  & \text{if} \quad r < R\,, \\[3mm]
\displaystyle \frac{Z_1 Z_2e^2}{r} & \text{if} \quad r \ge R\, ,
\end{cases}
\end{equation}
where $Z_1 = 2$ and $Z_2 = Z-2$ in our case.
The radius $R$ of the core nucleus is found from the Bohr-Sommerfeld quantization condition:
\begin{equation}
\int_{r_1^{}}^{r_2^{}} dr\, k(r) = \left(n + \textstyle\frac{1}{2} \right) \pi
= (G - \ell + 1)\frac{\pi}{2}\, ,
\label{eq:BS_quant}
\end{equation}
where the value of $G$ depends on the neutron number $N$ as~\cite{BMP91, BMP92}
\begin{equation}
G = \left\{ \begin{array}{ll}
20 \quad & \mbox{for } N \le 82, \\
22 \quad & \mbox{for } 82 < N \le 126, \\
24 \quad & \mbox{for } 126 < N. \end{array} \right.
\end{equation}

The centrifugal barrier is written as
\begin{equation}
V_L = \frac{\hbar^2}{2m_\mu^{}r^2} \ell (\ell + 1) ,
\end{equation}
where $\ell$ is the relative orbital angular momentum between the core nucleus and the $\alpha$
particle.
In this calculation, $\ell=0$ is assumed as in Refs.~\cite{BMP91, BMP92} and thus there is no 
contribution from $V_L$.%
\footnote{However, as will be discussed in Sec.~\ref{sec:results}, the assumption of $\ell=0$ is 
too crude and gives reasonable results only for even-even nuclei.}

On top of this model, we consider the effects of the nuclear symmetry energy.
In order to take into account the isospin asymmetry effects, we modify the $\alpha$ particle 
nuclear potential for $r < R$ as
\begin{equation}
V_N(r) = V_0 + V_1 I + V_2 I^2,
\end{equation}
where 
\begin{equation}
I  = (N-Z)/A = (N-Z)/(N+Z)
\label{eq:definition_I}
\end{equation}
with $Z$ being the number of protons so that $A = N+Z$.
The constants $V_1$ and $V_2$ control the dependence of the nuclear potential
on the isospin asymmetry.

\subsection{\boldmath Woods-Saxon potential for $\alpha$ particle}

More realistic potentials than the simplest square well potential can be constructed by considering
nonuniform distribution of nucleons in the core nucleus.
One typical example is the Woods-Saxon potential~\cite{WS54} which assumes 
Fermi or logistic function distribution of the nucleon density profile.
This leads to the nuclear potential in the Woods-Saxon form:
\begin{equation}
V_N(r) = \frac{V_0}{1+\exp\left[(r-R)/a \right] }\,,
\end{equation}
where $R$ is the rough radius of the nucleus and $a$ is diffuseness parameter. 
As in the case of the square well potential model in the previous subsection, the radius $R$ is 
determined by the quantization condition of Eq.~(\ref{eq:BS_quant}).
To take into account isospin asymmetry, we modify $V_N(r)$ as
\cite{DI05, LZZS10, VTMLC91}
\footnote{
We thank the referee for pointing out
that the nuclear potential may also include the linear term of $I$.
}
\begin{equation}
V_N(r) = \frac{1}{1+\exp\left[(r-R)/a \right] } \left( V_0 + V_1 I + V_2 I^2 \right).
\end{equation}
The value of $a$ obtained from the least $\sigma$ fitting is found to be $a=0.4$~fm.%
\footnote{
In Fermi density distribution, the surface diffuseness $a$ is roughly $t_{90 \mbox{-} 10}/4.4$, 
where $t_{90 \mbox{-} 10}$ is the radial distance between 90\% and 10\% of the density peak.
For example, $t_{90\mbox{-}10}$ thickness for \nuclide[208]{Pb} from SLy4 Hartree-Fock calculation 
is about $2.60$~fm, which then gives $a \approx 0.59$~fm~\cite{CBHMS98}. 
This value differs from the fitted value by a factor of $1.5$. }
We also tried to improve the fitting by allowing the functional form of $a = a(I)$, but it does 
not show any apparent isospin dependence in minimizing the rms deviation. 
Thus, we fix $a=0.4$~fm in our simulation as in Ref.~\cite{BMP92b}.

In this model, the core nucleus is assumed to have a uniform charge distribution~\cite{BMP92b}.
Therefore, unlike the square well potential model, we have
\begin{equation}
V_C(r) = 
\begin{cases} 
\displaystyle \frac{Z_1Z_2 e^2}{2R} \left[3 - \left(\frac{r}{R}\right)^2 \right]
& \text{for }\quad r < R\, ,  \\[3mm]
\displaystyle \frac{Z_1 Z_2e^2}{r}       & \text{for } \quad  r \ge R\, .
\end{cases}
\end{equation}
Furthermore, it is known that the proper application of the WKB formula needs to replace $V_L$ by
the modified centrifugal barrier of Langer~\cite{Langer37}, which reads
\begin{equation}
V_L(r) = \frac{\hbar^2}{2m_\mu^{}r^2} \left( \ell + \frac12 \right)^2. 
\label{eq:Langer}
\end{equation}
In the present study, we vary the angular momentum in the effective $\alpha$ potential to
obtain the best fit with the experimental data, but with the constraint of parity conservation. 
This completes our second model for the $\alpha$ potential and the parameters are determined by 
minimizing the rms deviation defined in Eq.~(\ref{eq:rms}).

\subsection{\boldmath Potential based on the Skyrme energy density functional}

While the previous two models are based on macroscopic approaches to the nuclear potential of
the $\alpha$ particle, the phenomenological Skyrme force model gives a tool based on a more 
microscopic background to understand the form of the nuclear potential.
Within this approach the potentials of protons and neutrons in nuclei are expressed as functions of 
proton and neutron densities~\cite{VB72}.

As in the previous models, we assume that the $\alpha$ particle is small enough to be 
treated as a pointlike particle.
Then a pointlike $\alpha$ particle in a decaying nucleus interacts with pointlike nucleons in the core nucleus.
At the leading-order approximation, two-body interactions describe the interactions between the
$\alpha$ particle and the nucleons of the core nucleus.
Employing the standard form of the energy density functional (EDF) of the Skyrme force, 
we write the interaction of $\alpha$ particle as
\begin{eqnarray}
\label{eq:skyint}
v_{N\alpha}^{} (\bm{k},\bm{k}') &=& s_0^{} \left( 1+ v_0^{} P_\sigma \right) 
\delta ( \bm{r}_{N\alpha}^{})
\nonumber \\ && \mbox{} + \frac{s_1^{}}{2} \left( 1+ v_1^{} P_\sigma \right) 
\left[ \delta(\bm{r}_{N\alpha}^{}) \bm{k}^2 
+ \bm{k}'^2 \delta(\bm{r}_{N\alpha}^{}) \right]  
\nonumber \\ && \mbox{}
+ s_2^{} \, \bm{k}' \cdot \delta(\bm{r}_{N\alpha}^{}) \bm{k}  
\nonumber \\ && \mbox{}
+ i W_0^{\alpha} \bm{k}'  \cdot \left( \bm{\sigma} \times \bm{k} \right) 
\delta(\bm{r}_{N\alpha}^{}) 
\nonumber \\ && \mbox{}
+ \frac{s_3^{}}{6} \left( 1+v_3^{} P_\sigma \right) \rho_N^{\epsilon} 
\delta(\bm{r}_{N\alpha}^{})  \,,
\end{eqnarray}
where $\bm{r}_{N\alpha}^{} = \bm{r}_N^{} - \bm{r}_\alpha^{}$,
$\rho_N^{} = \rho_n^{} + \rho_p^{}$ is the nucleon density, 
$P_\sigma$ is the spin exchange operator, and $s_i^{}$, $v_i^{}$, $W_{0}^{\alpha}$ and $\epsilon$
are the parameters of the potential. 
The momenta $\bm{k}$ and $\bm{k}'$ are defined as%
\footnote{As usual, it is understood that $\bm{k}'$ operates to the left bra-space, while $\bm{k}$
operates to the right ket-space.}
\begin{equation}
\bm{k} = \frac{1}{2i} \left( \bm{\nabla}_{N}^{} - \bm{\nabla}_\alpha^{} \right),   \quad
\bm{k}' = -\frac{1}{2i} \left( \bm{\nabla}'_{N}-\bm{\nabla}'_\alpha \right) .
\end{equation}
Evaluating the matrix elements of Eq.~(\ref{eq:skyint}) leads to a form of the $\alpha$ particle potential 
as a functional of the proton and neutron densities as
\begin{eqnarray}
V_{N} & = & \alpha \rho_N^{}
+ \beta \left( \rho_n^{5/3} + \rho_p^{5/3} \right)
 + \gamma \rho^\epsilon_N  \left( \rho^2_N + 2 \rho_n^{} \rho_p^{} \right)
 \nonumber \\ && \mbox{}
 + \delta \frac{\rho_N^\prime}{r} + \eta \rho_N^{\prime\prime}\,,
\label{eq:VN_model}
\end{eqnarray}
where $\rho'_N = d \rho_N^{} / d r$ and $\rho''_N = d^2 \rho_N^{} / dr^2$.
Details for the derivation of Eq.~(\ref{eq:VN_model}) are described in Appendix.

As for the density profiles of protons and neutrons, we assume the Fermi distribution, i.e., the
form of the logistic function as
\begin{eqnarray}
\rho_n^{} &=& \frac{\rho_{n}^0}{1 + \exp\left[ \left(r - R_n \right)/a_n^{} \right] }\, ,
\nonumber \\
\rho_p^{} &=& \frac{\rho_{p}^0}{1 + \exp\left[ \left(r - R_p \right)/a_p^{} \right]}\, ,
\label{eq:denspro}
\end{eqnarray}
where $R_n$, $R_p$ are to be determined not from the quantization condition but from the number 
of neutrons and protons in the core nucleus. 
We use the values of $\rho_{n}^0$, $\rho_{p}^0$, $a_n$, and $a_p$
from the Thomas-Fermi calculation using the SLy4 force. 
Since the proton distribution is given explicitly, the Coulomb potential can be calculated as 
\begin{equation}
\begin{split}
V_C(r) & = 4\pi Z_1 e^2 \left[
\frac{1}{r}\int_0^r r^{\prime 2}\rho_p(r^\prime)\,dr^\prime
+ \int_{r}^{\infty}r^\prime \rho_p(r^\prime)dr^\prime\right]\,,
\end{split}
\end{equation}
where $Z_1 =2 $ in the case of $\alpha$ decay.

In the effective potential of Eq.~(\ref{eq:VN_model}), the isospin asymmetry effects are accounted 
for through the $\beta$ and $\gamma$ interaction terms.
The parameter $\epsilon$ is introduced to account for the nuclear many-body effects in nuclei, but
we found that the results are not sensitive to the value of $\epsilon$, so we set 
$\epsilon = \frac16$ throughout this study.
The interaction parameters $\alpha$, $\beta$, $\gamma$, $\delta$, and $\eta$ are fitted by 
minimizing the rms deviation. 
In this model, we also use the centrifugal barrier as given in Eq.~(\ref{eq:Langer}), and this completes
our model for $\alpha$ nuclear potential based on the Skyrme EDF.
\begin{table*}[t]
\caption{Parameters of the SW potential fitted to the experimental data of
Refs.~\cite{AWWK12,WAWK12}.
The numbers in parentheses denote the fitted values without the $V_1$  and $V_2$ terms. 
The rms deviation $\sigma$ is defined in Eq.~(\ref{eq:rms}).}
\begin{tabular}{c|ccccc}
\hline\hline
 Type & Number of events  & $V_0$ (MeV) & $V_1$ (MeV)
      &$V_2$ (MeV) & $\sigma$  \\
\hline
e-e   &  178  & $-140.035$ ($-132.415$) & $+57.567$ & $-71.601$  & 0.304 (0.319)  \\
e-o   &  110  & $-175.980$ ($-140.416$) & $+524.995$ & $-1737.533$  & 0.596 (0.616)  \\
o-e   &  137  & $-158.767$ ($-142.700$) & $+308.787$ & $-1163.721$  & 0.607 (0.630) \\
o-o   &  70   & $-152.100$ ($-144.250$) & $+56.482$ & $-63.256$  & 0.604 (0.609)\\
\hline\hline
\end{tabular}
\label{tb:squarei}
\end{table*}
\begin{table*}[t]
\caption{Parameters of the SW potential fitted to the experimental data of
Refs.~\cite{AWWK12,WAWK12}.
The numbers in parentheses denote the fitted values without the $V_1$ term. 
The rms deviation $\sigma$ is defined in Eq.~(\ref{eq:rms}).}
\begin{tabular}{c|cccc}
\hline\hline
 Type & Number of events  & $V_0$ (MeV) & $V_1$ (MeV) & $\sigma$  \\
\hline
e-e   &  178          & $-138.523$   ($-132.415$) & $+35.644$  & 0.304 (0.319)  \\
e-o   &  110          & $-135.823$   ($-140.416$) & $+25.727$  & 0.614 (0.616)  \\
o-e   &  137          & $-134.579$   ($-142.700$) & $-46.412$  & 0.620 (0.630) \\
o-o   &  70           & $-150.740$   ($-144.250$) & $+37.035$  & 0.604 (0.609)\\
\hline\hline
\end{tabular}
\label{tb:squareii}
\end{table*}
\begin{table*}[t]
\caption{Parameters of the SW potential fitted to the experimental data of
Refs.~\cite{AWWK12,WAWK12}.
The numbers in parentheses denote the fitted values without the $V_2$ term. 
The rms deviation $\sigma$ is defined in Eq.~(\ref{eq:rms}).}
\begin{tabular}{c|cccc}
\hline\hline
 Type & Number of events  & $V_0$ (MeV) & $V_2$ (MeV) & $\sigma$  \\
\hline
e-e   &  178          & $-135.933$   ($-132.415$) & $+111.431$  & 0.305 (0.319)  \\
e-o   &  110          & $-136.899$   ($-140.416$) & $-105.036$  & 0.612 (0.616)  \\
o-e   &  137          & $-136.969$   ($-142.700$) & $-175.735$  & 0.616 (0.630) \\
o-o   &  70           & $-148.022$   ($-144.250$) & $+116.513$  & 0.604 (0.609)\\
\hline\hline
\end{tabular}
\label{tb:squareiii}
\end{table*}

\begin{table}[t]
\caption{Fitted parameters of the WS potential. Notation is the same as in Table~\ref{tb:squarei}.}
\begin{tabular}{c|cccc}
\hline\hline
Type & $V_0$ (MeV) & $V_1$ (MeV) & $V_2$ (MeV) & $\sigma$ \\
\hline
e-e  & $-190.845$ ($-179.634$ ) & $+54.851$ & $+56.370$   & $0.302$ ($0.326$ ) \\
e-o  & $-173.564$ ($-174.859$ ) & $+64.534$ & $-38.600$   & $0.211$ ($0.212$) \\
o-e  & $-187.018$ ($-182.313$ ) & $+36.494$ & $+127.714$  & $0.248$ ($0.251$) \\
o-o  & $-180.316$ ($-176.876$ ) & $-16.653$ & $+86.544$   & $0.254$ ($0.256$) \\
\hline\hline
\end{tabular}
\label{tb:wsi}
\end{table}

\begin{table}[t]
\caption{Fitted parameters of the WS potential. Notation is the same as in Table~\ref{tb:squarei}.}
\begin{tabular}{c|ccc}
\hline\hline
Type & $V_0$ (MeV) & $V_1$ (MeV) & $\sigma$ \\
\hline
e-e  & $-191.785$ ($-179.634$ ) & $+70.737$  & $0.302$ ($0.326$ ) \\
e-o  & $-174.860$ ($-174.859$ ) & $-2.514$  & $0.212$ ($0.212$) \\
o-e  & $-182.293$ ($-182.313$ ) & $+0.965$  & $0.251$ ($0.251$) \\
o-o  & $-176.844$ ($-176.876$ ) & $-5.644$  & $0.256$  ($0.256$) \\
\hline\hline
\end{tabular}
\label{tb:wsii}
\end{table}

\begin{table}[t]
\caption{Fitted parameters of the WS potential. Notation is the same as in Table~\ref{tb:squarei}.}
\begin{tabular}{c|ccc}
\hline\hline
Type & $V_0$ (MeV) & $V_2$ (MeV) & $\sigma$ \\
\hline
e-e  & $-182.314$ ($-179.634$ ) & $+90.277$   & $0.311$ ($0.326$ ) \\
e-o  & $-173.971$ ($-174.859$ ) & $-22.157$  & $0.211$ ($0.212$) \\
o-e  & $-184.537$ ($-182.313$ ) & $+61.693$  & $0.249$ ($0.251$) \\
o-o  & $-178.532$ ($-176.876$ ) & $+42.256$  & $0.255$  ($0.256$) \\
\hline\hline
\end{tabular}
\label{tb:wsiii}
\end{table}

\subsection{\boldmath Empirical formula for $\alpha$ decay half-lives with isospin effects}

The Geiger-Nutall law gives a simple relationship of $\alpha$ decay lifetimes to the proton number and
the $Q_\alpha$ value~\cite{GN11}.  
The Viola-Seaborg (VS) empirical formula, which is an improved form of the Geiger-Nutall law, is widely 
used to estimate the $\alpha$ decay lifetimes, and it reads~\cite{VS66b}
\begin{equation}
\log_{10}(T_{1/2}/\text{s}) = \frac{aZ + b}{\sqrt{Q_\alpha/\text{MeV}}} + cZ + d \,,
\label{eq:VS}
\end{equation}
where $a$, $b$, $c$, $d$ are parameters to be fitted to the experimental data. 
In its original form, Eq.~(\ref{eq:VS}) contains an $h_{\rm log}$ term that takes into account the
difference between the even and odd nuclei.
In this work, we allow different values of the parameters for even or odd numbers of $Z$ and $N$, 
so introducing the blocking factor for odd nucleus $h_{\rm log}$ is not necessary in our formula.

The subsequent efforts to improve this relation can be found, e.g., 
in Refs.~\cite{RXW04,RZ08,NRDX08,SPC89,SSP15}.
Since the primary aim of the present work is to look for the effects of nuclear isospin asymmetry,
we simply modify the above formula as
\begin{equation}
\log_{10}(T_{1/2}/\text{s}) = \frac{aZ + b}{\sqrt{Q_\alpha/\text{MeV}}} + cZ + d +e_1I + e_2 I^2 ,
\end{equation}
where $I$ is defined in Eq.~(\ref{eq:definition_I}).

\section{Results}   \label{sec:results}

In this section, we perform the fitting procedure described in Sec.~\ref{subsec:general}
and present the fitted parameters. 
We then compare our results with the available experimental data and give our predictions
on the $\alpha$ decay lifetimes of superheavy elements.

\subsection{\boldmath Fitted parameters}

We begin with the simple square well (SW) potential model whose fitted parameters are
presented in Table~\ref{tb:squarei}, \ref{tb:squareii}, and \ref{tb:squareiii} 
for four different cases of $\alpha$ decays, namely, 
even-even (e-e), even-odd (e-o), odd-even (o-e), and odd-odd (o-o), where the former refers to 
the neutron number and the latter to the proton number of the decaying nucleus.
For the fitting process, the \textsc{Ame2012} experimental data compiled in 
Refs.~\cite{AWWK12,WAWK12} are used.
Numbers in parentheses denote the values obtained without the $I$ and $I^2$ terms.
Comparing the rms deviations $\sigma$ for the cases with and without the isospin asymmetry terms, 
we notice a slight improvement due to the $I$ and $I^2$ terms.
The rms deviation $\sigma$ value has the lowest value for the case of e-e nuclei and larger values 
for other nuclei.
The main reason for this behavior is the assumed value ($\ell = 0$) of the orbital angular momentum.
To verify this, we allow the variation of $\ell$ for each nuclei.
It is then found that $\ell = 0$ gives a reasonable description of the decays of even-even nuclei but
$\ell \neq 0$ is definitely needed to have a better fit for the other nuclei.
Since $\ell = 0$ is assumed for all nuclei in the SW potential model, there is a limit to reduce the 
$\sigma$ values for even-odd, odd-even, and odd-odd nuclei.
But we do not further pursue to find a better parameter set by varying the value of $\ell$ in this model, as
our purpose is to see the role of the isospin asymmetry terms
compared with the model of Refs.~\cite{BMP91,BMP92}.

Unlike the SW potential model discussed above, the $\ell = 0$ constraint is released in the 
Woods-Saxon (WS) potential model following the prescription of Ref.~\cite{BMP92b}.
Table~\ref{tb:wsi}, \ref{tb:wsii}, and  \ref{tb:wsiii} show the fitted parameters of the WS potential.
We can see that the rms deviation in the case of even-even nuclei is similar in quality to that 
of the SW potential model.
But the results for the other nuclei are improved a lot. 
The main reason is that, as was mentioned above, we allow the variation of $\ell$ in the fitting process.
Namely, we change the $\ell$ value for each nucleus so that it reproduces the best result of rms deviation, 
while the condition of parity conservation is satisfied. 
As a result, we obtain a better result for the rms deviation.
Inclusion of isospin asymmetry term slightly improves the results of even-even nuclei,
but leaves the rms deviation almost unchanged for odd-$N$ or odd-$Z$ nuclei, which implies
that, in this model, the angular momentum effect is much stronger than the isospin asymmetry effect.

\begin{table*}[t]
\caption{Fitted parameters of the $\alpha$ particle potential model based on the Skyrme EDF.}
\begin{tabular}{ccccc}
\hline\hline
$\alpha$     & $\beta$      & $\gamma$ & $\delta$ & $\eta$ \\
(MeV fm$^{3}$) & (MeV fm$^{5}$) & (MeV fm$^{6 +3\epsilon}$) &
~(MeV fm$^{5}$)~ & ~(MeV fm$^{5}$)~ \\
\hline
~$-1.6740\times 10^{3} $~ &
~$ 1.9208\times 10^{3} $~ &
~$ 1.7182\times 10^{3} $~ &
~$ 9.4166 $~ &
~$ -26.7616$~\\
\hline\hline
$\sigma(\mbox{e-e})$ & $\sigma(\mbox{e-o})$ & $\sigma(\mbox{o-e}$) & $\sigma(\mbox{o-o})$ & 
$\sigma(\mbox{All})$ \\
\hline
\ 0.319 \    & \ 0.276 \   & \ 0.283 \   & \ 0.301 \  & \ 0.296 \    \\
\hline\hline
\end{tabular}\label{tb:EDF_pam}
\end{table*}

Presented in Table~\ref{tb:EDF_pam} are the parameters of the $\alpha$ nuclear potential from the 
Skyrme EDF.
For the fitting process, we use the data only for even-even nuclei since the formula already includes the 
isospin dependence explicitly and the parameters should be the same for the four cases of the proton
and neutron numbers.
Table~\ref{tb:EDF_pam} also displays the rms deviation values with the fitted parameters
for four cases of nuclei.
As in the WS potential model, we assume $\ell = 0$ for even-even nuclei, but allow the change of 
$\ell$ for other nuclei, which results in smaller deviations for odd-$N$ or odd-$Z$ nuclei.
The overall agreement with the measured data is as satisfactory as the WS potential model.
More detailed comparison will be presented in the next subsection.


\begin{figure}[t]
\centering
\includegraphics[width = 0.95\columnwidth]{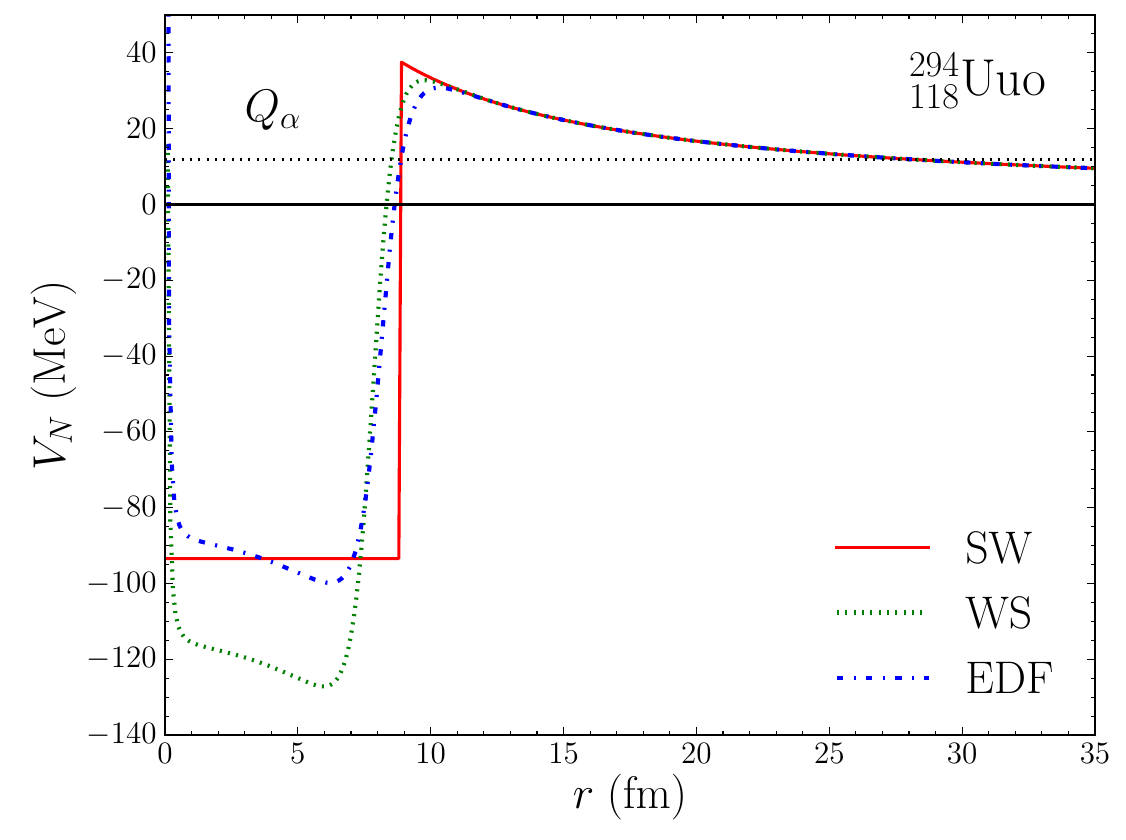}
\caption{(Color online)
The $\alpha$ particle nuclear potential in the nucleus \nuclide[294][118]{Uuo} of the three models
considered in the present work.
}
\label{fig:pot}
\end{figure}

In order to see the model dependence of the results, we plot the obtained nuclear potentials of
the $\alpha$ particle for the nucleus \nuclide[294][118]{Uuo} in Fig.~\ref{fig:pot}. 
We find that the three models provide similar potentials but the structure of the potential in the inner 
region ($r <10$~fm) shows rather strong model dependence.
Namely, the WS potential gives the deepest potential, while the depths of SW and EDF potentials 
are similar to each other.
Roughly speaking, the depth of the WS potential is bigger than those of the SW and EDF potentials 
by about 20\%.
On the other hand, the barrier width for a given value of $Q_\alpha$ takes the largest value for 
the WS potential and the smallest for the EDF one, but the difference is only less than 1~fm.
Since the half-life is mostly determined by the quantum tunneling effects, the major factor that
determines the lifetime is the potential width where the $\alpha$ particle should penetrate.
Therefore, in the case of \nuclide[294][118]{Uuo}, we have the hierarchy of
$T^{\rm SW}_{1/2} \simeq T^{\rm WS}_{1/2}>T^{\rm EDF}_{1/2}$ that is confirmed by
numerical calculation.%
\footnote{Of course, we have different relations among them depending on the nucleus.} 
This is so because a shorter penetration barrier gives a shorter lifetime.
However, the inner part of the potential may affect the lifetime through the assaulting frequency 
$\mathcal{F}$ determined by the $Q_\alpha$ value. 
The results shown in Tables \ref{tb:squarei} --\ref{tb:EDF_pam} suggest that the isospin asymmetry 
effects in SW and WS models are mostly involved in assaulting frequencies and the penetration lengths 
are almost unaffected. 
Therefore, the rms deviations are not improved much by the isospin asymmetry effect.

\begin{table*}[t]
\caption{
Fitted coefficients of the modified VS formula. 
The values in parentheses are those of the unmodified VS formula, i.e., without the $e_1$ and $e_2$ terms.}
\begin{tabular}{c|ccccccc}
\hline\hline
Type & $a$ & $b$ & $c$ & $d$ &  $e_1$ & $e_2 $ & $\sigma$ \\
\hline
e-e & $1.53420$  ($1.48503$)        & $4.20759$ ($5.26806$) & $-0.18124$ ($-0.18879$) 
    & $-35.57934$ ($-33.89407$ )    & $5.28401$ & $-38.17144$ & $0.311 $ $(0.359)$    \\  
e-o &  $1.64322$ ($1.55427$)    &  $-2.33315$ ($1.23165$) &  $-0.18749$ ($-0.18838$) 
    &  $-35.27841$  ($-34.29805$ ) &$1.19898$   & $-31.24030$  & $0.571 $ $(0.608)$   \\  
o-e &  $1.69868  $   ($1.64654$)     &  $-5.67266$ ($-3.14939$) &  $-0.22366$ ($-0.22053$) 
    &  $-32.02953$ ($-32.74153$)  &$-12.96399$  & $31.01813$   & $0.542 $ $(0.554)$      \\  
o-o &  $1.37778$ ($1.34355$)        &  $13.63138$ ($13.92103$) &  $-0.11009$ ($-0.12867$) 
    &  $-39.41075$ ($-37.19944$)   & $5.98423$ & $-52.56801     $   &$0.561 $ $(0.617)$  \\  
\hline\hline
\end{tabular}\label{tb:vsformi}
\end{table*}

\begin{table*}[t]
\caption{
Fitted coefficients of the modified VS formula. 
The values in parentheses are those of the unmodified VS formula, i.e., without the $e_1$ term.}
\begin{tabular}{c|cccccc}
\hline\hline
Type & $a$ & $b$ & $c$ & $d$ &  $e_1 $ & $\sigma$ \\
\hline
e-e & $1.53223$  ($1.48503$)        & $4.33481$ ($5.26806$) & $-0.18002$ ($-0.18879$) 
    & $-34.97023$ ($-33.89407$ )   & $-5.77017$ & $0.327 $ $(0.359)$    \\  
e-o &  $1.62853$     ($1.55427$)    &  $-1.43833$ ($1.23165$) &  $-0.18372$ ($-0.18838$) 
    &  $-34.83250$  ($-34.29805$ ) & $ -8.12715$   &$0.573 $ $(0.608)$   \\ 
o-e &  $1.68262 $   ($1.64654$)     &  $-4.30929$ ($-3.14939$) &  $-0.21807$ ($-0.22053$) 
    &  $-33.11939$ ($-32.74153$)    & $-3.73747$   & $0.547 $ $(0.554)$      \\  
o-o &  $1.43614$ ($1.34355$)        &  $10.05247$ ($13.92103$) &  $-0.12664$ ($-0.12867$) 
    &  $-37.45332$ ($-37.19944$)    & $-9.64627$   & $0.570 $ $(0.617)$  \\  
\hline\hline
\end{tabular}\label{tb:vsformii}
\end{table*}

\begin{table*}[t]
\caption{
Fitted coefficients of the modified VS formula. 
The values in parentheses are those of the unmodified VS formula, i.e., without the $e_2$ terms.}
\begin{tabular}{c|cccccc}
\hline\hline
Type & $a$ & $b$ & $c$ & $d$ &  $e_2 $ & $\sigma$ \\
\hline
e-e & $1.53829$  ($1.48503$)        & $4.16407$ ($5.26806$) & $-0.17981$ ($-0.18879$) 
    & $ -35.39178$ ($-33.89407$ )   & $-22.00448$ & $0.314 $ $(0.359)$    \\  
e-o &  $1.64186$     ($1.55427$)    &  $-2.23887$ ($1.23165$) &  $-0.18700$ ($-0.18838$) 
    &  $ -35.22664$  ($-34.29805$ ) & $ -27.38784$   & ~~$0.571 $ $(0.608)$   \\  
o-e &  $1.66663 $   ($1.64654$)     &  $-3.56210$ ($-3.14939$) &  $-0.21732$ ($-0.22053$) 
    &  $-33.29803$ ($-32.74153$)    & $-8.53409$   & ~~$0.550 $ $(0.554)$      \\  
o-o &  $1.40242$ ($1.34355$)        &  $12.19381$ ($13.92103$) &  $-0.11593$ ($-0.12867$) 
    &  $-38.72067$ ($-37.19944$)    & $-33.75875$   & ~~$0.563 $ $(0.617)$  \\  
\hline\hline
\end{tabular}\label{tb:vsformiii}
\end{table*}

In the present work, we also investigate the modified VS formula for the $\alpha$ decay lifetimes.
Table \ref{tb:vsformi}--\ref{tb:vsformiii} show our results on the fitted parameters for the modified VS formula and 
the corresponding rms deviation. 
The numbers in parentheses represent the results without the isospin asymmetric terms.
Compared to the SW and WS potential models, inclusion of the isospin asymmetry term considerably 
improves the rms deviation.
However, the obtained rms deviations are larger than the WS model which may indicate some missed
structure in the VS formula.
Firstly, in the (modified) VS formula, there is no room to incorporate the contribution of the angular 
momentum $\ell$, i.e., the centrifugal barrier, so this may limit the application of the VS formula.
Secondly, the $\alpha$ decay lifetimes may have a more complicated dependence on isospin asymmetry
other than the $I$ and $I^2$ terms.
Such effects could be accounted for through more realistic microscopic approaches.


\begin{table*}[t]
\caption{Results for $\alpha$ decay half-lives of heavy nuclei.
The upper and lower bounds of theoretical calculations are from the experimental errors of
$Q_\alpha$ values.
} \label{tb:heavy}

\begin{tabular}{c|ccccccccc}
\hline\hline
$(Z,A)$       &  $Q_{\alpha}^{\text{Expt.}}$ (MeV) & $T_{1/2}^{\text{Expt.}}$  
              & $T_{1/2}^{\text{SW}}$ & $T_{1/2}^{\rm WS}$
              & $T_{1/2}^{\text{EDF}}$ & $T_{1/2}^{\rm VS}$
& Ref.  \\
\hline
~$(118, 294)$~ &  ~$11.81 \pm 0.06$~  & ~$0.89_{-0.31}^{+1.07}$ ms~ 
& ~$1.46^{+0.51}_{-0.38}$ ms~  
& ~$1.26^{+0.45}_{-0.33}$ ms~  
& ~$0.40^{+0.15}_{-0.11}$ ms~  
& ~$0.31^{+0.12}_{-0.08}$ ms~  
& \cite{OULA06}              \\
$(116,293)$ &  $10.67 \pm 0.06$    & $53_{-19}^{+62}$ ms 
&  ~$163^{+69}_{-48}$ ms 
&  $104^{+44}_{-31}$ ms    
&  $52^{+23}_{-16}$ ms   
&  $181^{+84}_{-57}$ ms     
& \cite{OULA04b}            \\
$(116,292)$ &  $10.80 \pm 0.07$           & $18_{-6}^{+16}$ ms 
& ~$78^{+39}_{-26}$ ms~ 
& $69^{+35}_{-23}$ ms  
& $25^{+13}_{-8}$ ms  
& $20^{+10}_{-7}$ ms 
& \cite{OULA04b}   \\
$(116,291)$ &  $10.89 \pm 0.07$    & $6.3_{-2.5}^{+11.6}$ ms 
& $47^{+23}_{-15}$ ms     
& $31^{+15}_{-10}$ ms  
& $16^{+8}_{-5}$ ms    
& $46^{+25}_{-16}$ ms 
& \cite{OULA06}  \\
$(116,290)$ &  $11.00 \pm 0.08$           & $7.1_{-1.7}^{+3.2}$ ms
& ~$25.9^{+14.5}_{-9.2}$ ms~  
& $23.2^{+13.2}_{-8.3}$ ms   
& $8.9^{+5.0}_{-3.3}$ ms  
& $7.2^{+4.2}_{-2.6}$ ms
& \cite{OULA06}  \\
$(115,288)$ &  $10.61 \pm 0.06$    & $87_{-30}^{+105}$ ms 
& $115^{+48}_{-34}$ ms    
& $139^{+60}_{-41}$ ms  
& $43^{+19}_{-13}$ ms  
& $676^{+279}_{-196}$ ms 
& \cite{OULA04-OUDL05}  \\
$(115,287)$ &  $10.74 \pm 0.09$    & $32_{-14}^{+155}$ ms 
&~$55^{+37}_{-22}$ ms~       
& $50^{+34}_{-20}$ ms    
& $21^{+15}_{-8}$ ms   
& $131^{+97}_{-55}$ ms
& \cite{OULA04-OUDL05}  \\
$(114,289)$ &  $9.96 \pm 0.06$   & $2.7_{-0.7}^{+1.4}$ s 
& $2.8^{+1.3}_{-0.9}$ s    
& $3.1^{+1.5}_{-1.0}$ s   
& $1.1^{+0.5}_{-0.3}$ s  
& $4.8^{+2.5}_{-1.6}$ s   
& \cite{OULA04b}          \\
$(114,288)$ &  $10.09 \pm 0.07$    & $0.8_{-0.18}^{+0.32}$ s
 & ~$1.2^{+0.68}_{-0.43}$ s~  
 & $1.12^{+0.63}_{-0.40}$ s   
 & $0.48^{+0.27}_{-0.17}$ s  
 & $0.39^{+0.22}_{-0.14}$ s    
 & \cite{OULA04b}           \\
$(114,287)$ &  $10.16 \pm 0.06$   & $0.48_{-0.09}^{+0.16}$ s
& $0.80^{+0.36}_{-0.25}$ s 
& $0.53^{+0.24}_{-0.17}$ s   
& $0.32^{+0.15}_{-0.10}$ s  
& $1.23^{+0.61}_{-0.41}$ s   
& \cite{OULA06}         \\
$(114,286)$ &  $10.33 \pm 0.06$           & $0.13_{-0.02}^{+0.04}$ s
& ~$0.29^{+0.13}_{-0.09}$ s~ 
& $0.26^{+0.12}_{-0.08}$  s  
& $0.12^{+0.05}_{-0.04}$  s 
& $0.10^{+0.04}_{-0.03}$  s   
& \cite{OULA06}    \\
$(113,284)$ &  $10.15 \pm 0.06$  & $0.48_{-0.17}^{+0.58}$ s
& $0.40^{+0.18}_{-0.12}$ s  
& $0.50^{+0.23}_{-0.16}$ s  
& $0.28^{+0.13}_{-0.09}$ s  
& $2.12^{+0.93}_{-0.64}$ s        
& \cite{OULA04-OUDL05}      \\
$(113,283)$ &  $10.26 \pm 0.09$  & $100_{-45}^{+490}$ ms 
&~$209^{+152}_{-87}$ ms~    
& $62^{+45}_{-26}$ ms   
& $91^{+69}_{-39}$ ms  
& $563^{+445}_{-246}$ ms  
& \cite{OULA04-OUDL05}          \\
$(113,282)$ &  $10.83 \pm 0.08$   & $73_{-29}^{+134}$ ms 
& $8^{+4}_{-3}$ ms 
& $52^{+30}_{-19}$ ms   
& $75^{+44}_{-28}$ ms  
& $52^{+29}_{-18}$ ms
& \cite{OULA07}         \\
$(112,285)$ &  $9.29 \pm 0.06$            & $34_{-9}^{+17}$ s
& $50^{+27}_{-17}$ s 
& $34^{+18}_{-12}$ s   
& $23^{+12}_{-8}$ s  
& $133^{+76}_{-48}$ s      
& \cite{OULA04b}       \\
$(112,283)$ &  $9.67 \pm 0.06$    & $3.8_{-0.7}^{+1.2}$ s 
& $3.9^{+1.9}_{-1.3}$ s  
& $4.5^{+2.2}_{-1.5}$ s   
& $1.8^{+0.9}_{-0.6}$ s   
& $8.4^{+4.4}_{-2.9}$ s   
& \cite{OULA06}   \\
$(111,280)$ &  $9.87 \pm 0.06$   & $3.6_{-1.3}^{+4.3}$ s
& $0.50^{+0.23}_{-0.16}$ s 
& $3.7^{+1.7}_{-1.2}$ s   
& $6.0^{+2.9}_{-1.9}$ s  
& $2.4^{+1.1}_{-0.7}$ s    
& \cite{OULA04-OUDL05}              \\
$(111,279)$ &  $10.52 \pm 0.16$  & $170_{-80}^{+810}$ ms 
& ~$10^{+16}_{-6}$ ms ~      
& $62^{+96}_{-37}$ ms    
& $110^{+177}_{-67}$ ms   
& $23^{+39}_{-14}$ ms   
& \cite{OULA04-OUDL05}      \\
$(111,278)$ &  $10.89 \pm 0.08$   & $4.2_{-1.7}^{+7.5}$ ms 
& $1.4^{+0.7}_{-0.5}$ ms   
& $2.7^{+1.5}_{-0.9}$ ms     
& $2.7^{+1.6}_{-1.0}$ ms   
& $8.2^{+4.4}_{-2.9}$ ms        
& \cite{OULA07}     \\ 
$(110,279)$ &  $9.84 \pm 0.06$            & $0.20_{-0.04}^{+0.05}$ s
& $0.28^{+0.13}_{-0.09}$ s   
& $0.18^{+0.08}_{-0.06}$ s    
& $0.13^{+0.06}_{-0.04}$ s   
& $0.59^{+0.30}_{-0.20}$ s    
& \cite{OULA06}         \\
$(109,276)$ &  $9.85 \pm 0.06$    & $0.72_{-0.25}^{+0.97}$ s
& $0.12^{+0.05}_{-0.04}$ s  
& $0.88^{+0.41}_{-0.28}$ s    
& $0.29^{+0.14}_{-0.09}$ s  
& $0.52^{+0.23}_{-0.16}$ s    
& \cite{OULA04-OUDL05}         \\
$(109,275)$ &  $10.48 \pm 0.09$  & $9.7_{-4.4}^{+46}$ ms 
& $3.0^{+2.0}_{-1.2}$ ms   
& $18.6^{+12.5}_{-7.4}$ ms    
& $6.7^{+4.6}_{-2.7}$ ms   
& $6.3^{+4.5}_{-2.6}$ ms      
& \cite{OULA04-OUDL05}      \\
$(109,274)$ &  $9.95 \pm 0.10$  & $440_{-170}^{+810}$ ms 
& $67^{+56}_{-30}$ ms   
& $480^{+416}_{-220}$ ms   
& $172^{+153}_{-80}$ ms  
& $353^{+294}_{-159}$ ms 
& \cite{OULA07}  \\
$(108,275)$ &  $9.44 \pm 0.06$   & $0.19_{-0.07}^{+0.22}$ s
& $0.75^{+0.36}_{-0.24}$ s  
& $0.48^{+0.24}_{-0.16}$ s    
& $0.39^{+0.20}_{-0.13}$ s  
& $2.12^{+1.12}_{-0.73}$ s       
& \cite{OULA06}         \\
$(107,272)$ &  $9.15 \pm 0.06$            & $9.8_{-3.5}^{+11.7}$ s
& $2.3^{+1.2}_{-0.8}$ s  
& $5.3^{+2.8}_{-1.8}$ s     
& $7.0^{+3.7}_{-2.4}$ s   
& $8.7^{+4.3}_{-2.9}$ s  
& \cite{OULA04-OUDL05}             \\
$(107,270)$ &  $9.11 \pm 0.08$      & $61_{-28}^{+292}$ s 
& $3.1^{+2.3}_{-1.3}$ s  
& $25^{+19}_{-11}$ s   
& $60^{+46}_{-26}$ s  
& $14^{+10}_{-6}$ s     
& \cite{OULA07}     \\
$(106,271)$ &  $8.67 \pm 0.08$  & $1.9_{-0.6}^{+2.4}$ min
& ~$0.51^{+0.41}_{-0.22}$ min~  
& ~$2.06^{+1.71}_{-0.92}$ min~  
& ~$1.67^{+1.41}_{-0.76}$ min~  
& ~$2.28^{+2.01}_{-1.06}$ min~          
& \cite{OULA06}       \\
\hline
$\sigma$ &  - &  - & $0.616$ & $0.290$ & $0.238$ & $0.513$ & \\
\hline\hline
\end{tabular}
\end{table*}

\subsection{\boldmath Comparison with data}
\label{sec:heavy}

We present our results for $\alpha$ decay half-lives of several heavy nuclei in Table~\ref{tb:heavy},
which shows the results from the SW potential model, WS potential model, Skyrme EDF potential model,
and the VS formula, where the SW, WS, and VS models include the isospin asymmetry terms.
The experimental $Q_\alpha$ values and measured half-lives of heavy nuclei are also given 
for comparison.
The rms deviation $\sigma$ given in this table is the value obtained with the listed 27 nuclei.
All the models give half-lives consistent with the experimental data and, at least, they are in the correct
order of magnitude.
Very few exceptional cases are the SW and VS models for the $(Z,\, A) = (111,\, 279)$ nucleus
and SW model for the cases of $(107,\, 270)$ and $(109,\, 274)$, where the theoretical values 
are smaller than the measured data by an order of magnitude.
On the other hand, Skyrme EDF model reproduces the experiment data fairly well,
giving the ratio of theory to experiment in the range from 0.40 for $(109,\, 276)$
to 2.53 for $(116,\, 291)$.%
\footnote{Our fitted parameters determined in this section without the isospin term are
consistent with those of Refs.~\cite{BMP91,BMP92,SPC89} for even-even nuclei considering
the different sets of data used in the fitting procedure. 
In the present work, for the SW and WS potentials, we carry out the fitting
separately for even-even, even-odd, odd-even, and odd-odd nuclei.}

Excellence of the EDF approach for the listed 27 heavy nuclei can be verified by the small value
of the rms deviation as shown in the last row of Table~\ref{tb:heavy}.
For the SW potential and the VS formula, the $\sigma$ values are significantly larger than the 
values given in Tables~\ref{tb:squarei} and \ref{tb:wsi} that are obtained in the fitting.
This may indicate the limitation of these models to describe $\alpha$ decays of heavy nuclei.
As was mentioned earlier, the orbital angular momentum $\ell$ is set to be zero in the SW and VS models 
regardless of the type of decaying nuclei.
On the other hand, this restriction is released for the WS and Skyrme EDF models, and
consequently, they lead to better fittings.
Nevertheless, it should be mentioned that the rms deviation value of the Skyrme EDF model in 
Table~\ref{tb:heavy} is even smaller than those in Table~\ref{tb:EDF_pam} and this
indicates the usefulness of this model for describing $\alpha$ decays of heavy nuclei.

We also compare our results with those obtained in the unified fission model (UFM) of 
Ref.~\cite{DZGWP10}.
For this end we take Table~I in Ref.~\cite{DZGWP10} as a benchmark for our calculation to have 
one-to-one comparison possible.
For most nuclei UFM also gives a good agreement with the experiment data except for several cases 
such as $(Z,\, A)=$ (107, 270), (109, 274), (111, 279), and (113, 282), where the UFM predictions
are smaller than the measured data by a factor of 10 or more.
On the other hand, the Skyrme EDF model of the present work gives quite reasonable description 
for these cases.
The main reason is attributed to the fact that, in the Skyrme EDF model, the shape of the potential 
changes depending on the values of $Z$ and $A$.
By changing the values of $Z$ and/or $A$, the parameters of 
the density profiles of protons and neutrons 
in Eq.~(\ref{eq:denspro}) need to be re-adjusted to find the minimum energy condition, 
which leads to the modification of the potential and thus the penetration length.
Although the SW and WS potentials are somehow dependent on the neutron number through 
the quantization condition of Eq.~(\ref{eq:BS_quant}),%
\footnote{A part of the isospin asymmetry effects of $\alpha$ decay comes from the penetration length 
of the Coulomb potential which depends only on $Z$.}
the resulting half-lives indicate that the EDF model treats the modification of potential in a more 
proper way.
This again suggests that microscopic treatments of nuclear potential are needed for more  
realistic approaches for understanding nuclear phenomena.

\subsection{\boldmath Predictions on undiscovered $\alpha$ decay lifetimes of superheavy elements}
\label{sec:prediction}

The information on the $\alpha$ decay lifetime can help experimentally confirm the synthesis of 
unknown superheavy elements.
In this subsection we present our predictions on $\alpha$ decays of such elements.
In this case, however, we do not have reliable information on the value of $Q_\alpha$, so we have to 
rely on the predictions of theoretical models on nuclear structure.
Since the $\alpha$-decay lifetime is sensitive to the value of $Q_\alpha$, this causes uncertainties in
our estimates.
In our calculation, we use the recent Weizs\"{a}cker-Skyrme4 (WS4) model of Ref.~\cite{WLWM14},
which gives a good description for the nuclei of $Z \ge 100$.
(See, for example, Refs.~\cite{MNMS93,DZ94,GCP10} for other models.)
With nuclei masses the $Q_\alpha$ values can be calculated by~\cite{MRDT06}
\begin{eqnarray}
Q_\alpha &=&  \Delta M(Z,A) - \Delta M(Z-2, A-4) - \Delta M_{\alpha}
\nonumber \\ && \mbox{}
+ 10^{-6}\, k\left[ Z^\beta - (Z-2)^\beta \right] ,
\end{eqnarray}
where $\Delta M$ is the atomic mass-excess,
$\Delta M_\alpha = 2.4249$ MeV, and ($k=8.7$ eV, $\beta =2.517 $) for nuclei of $Z \ge 60$ and 
($k=13.6$ eV, $\beta =2.408 $) for nuclei of $Z < 60$.

\begin{figure*}[t]
\centering
\includegraphics[width=0.8\columnwidth]{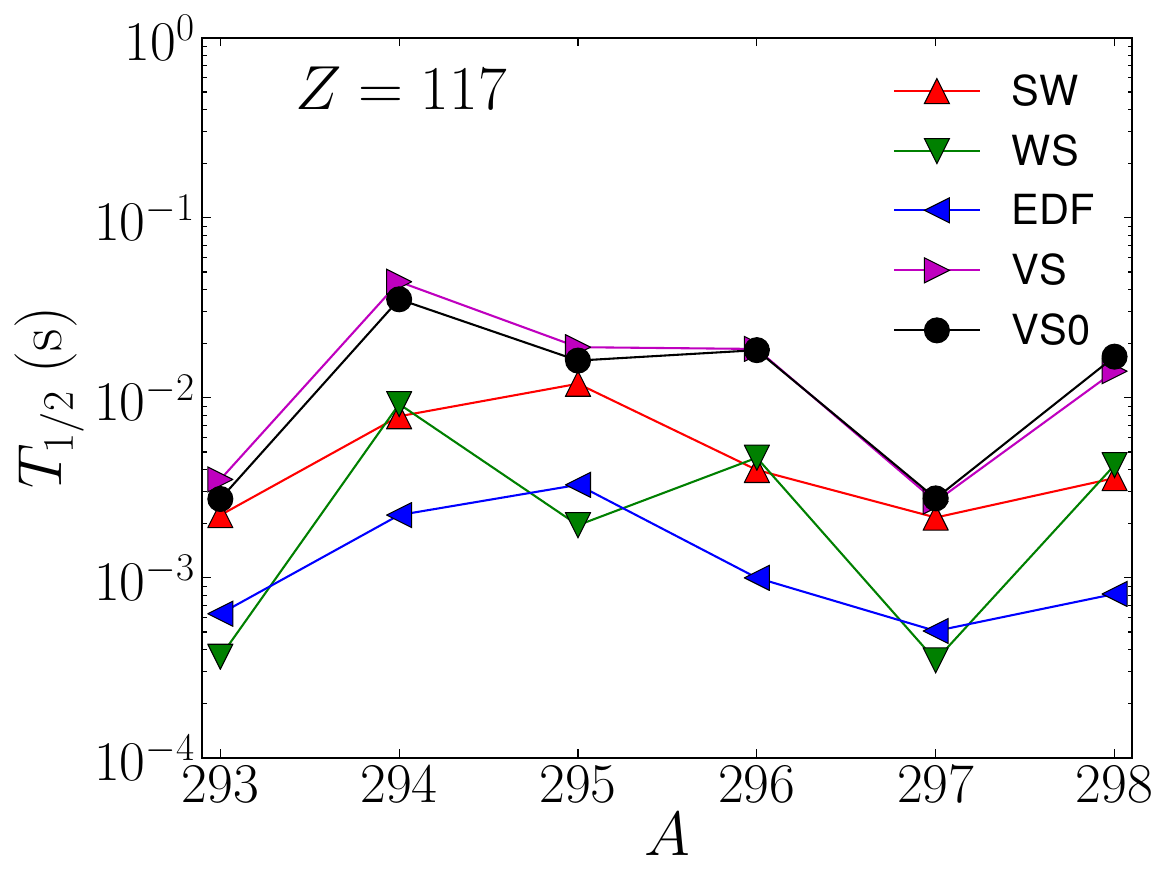}
\includegraphics[width=0.8\columnwidth]{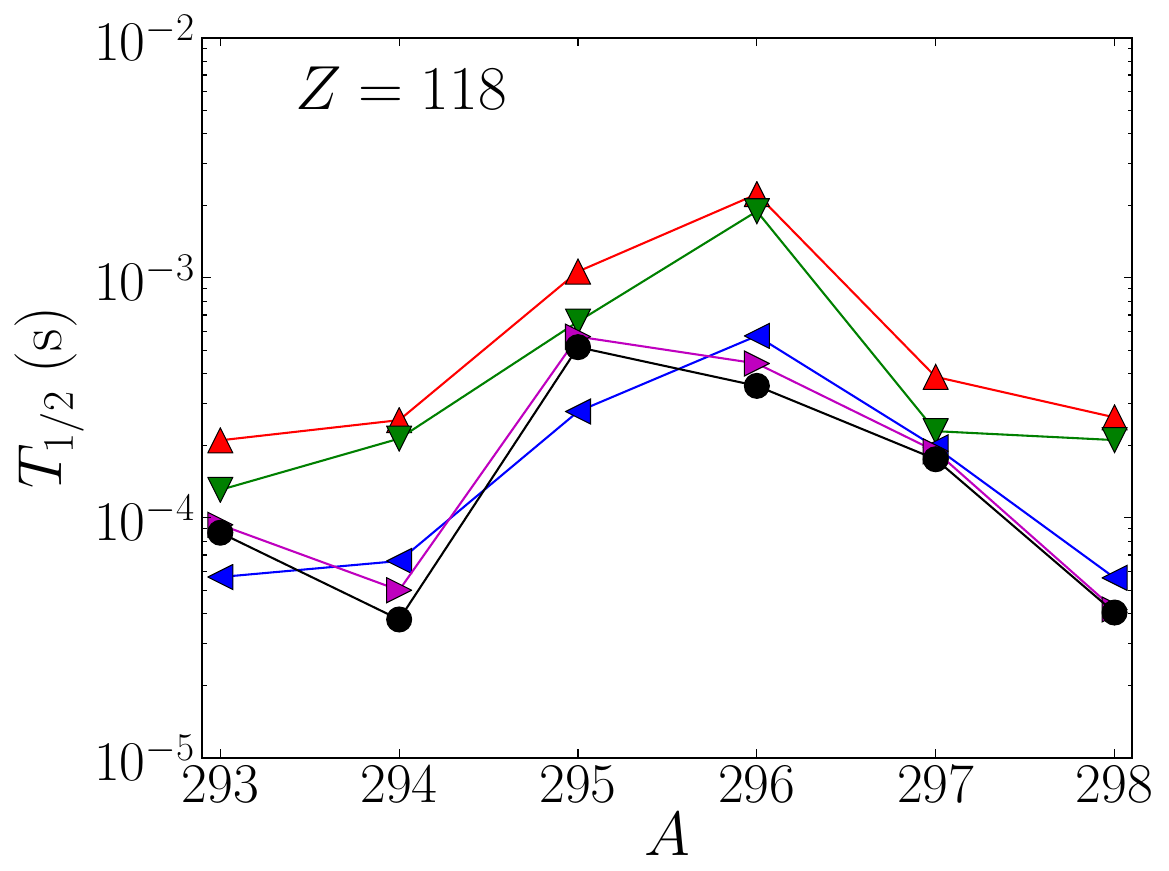}
\includegraphics[width=0.8\columnwidth]{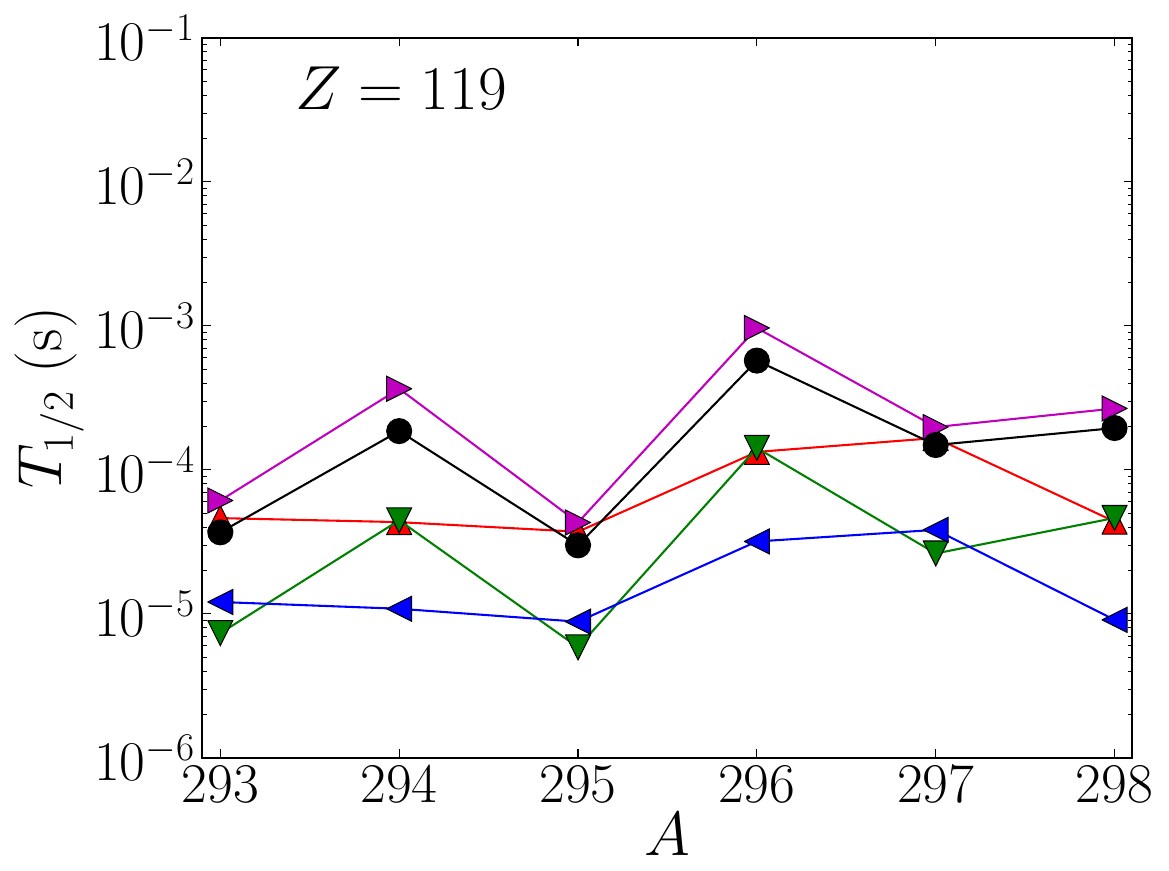}
\includegraphics[width=0.8\columnwidth]{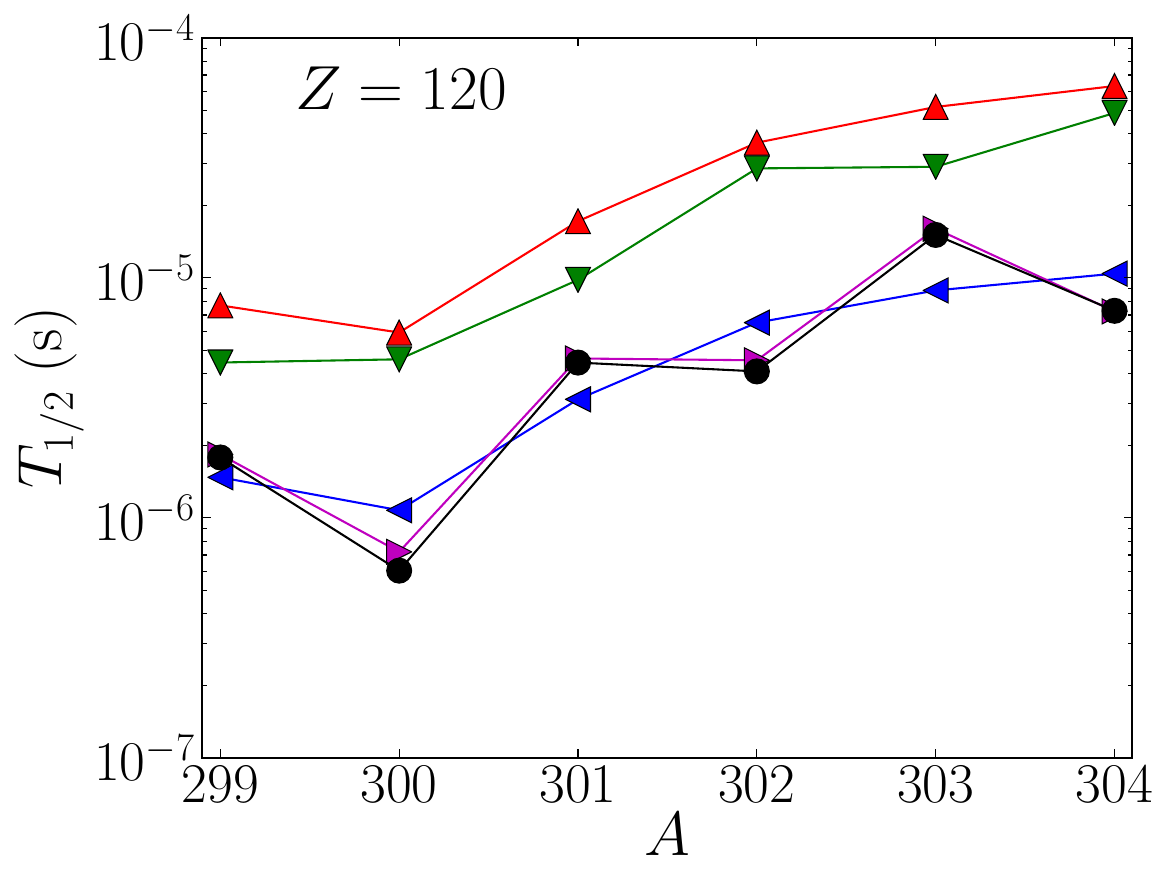}
\includegraphics[width=0.8\columnwidth]{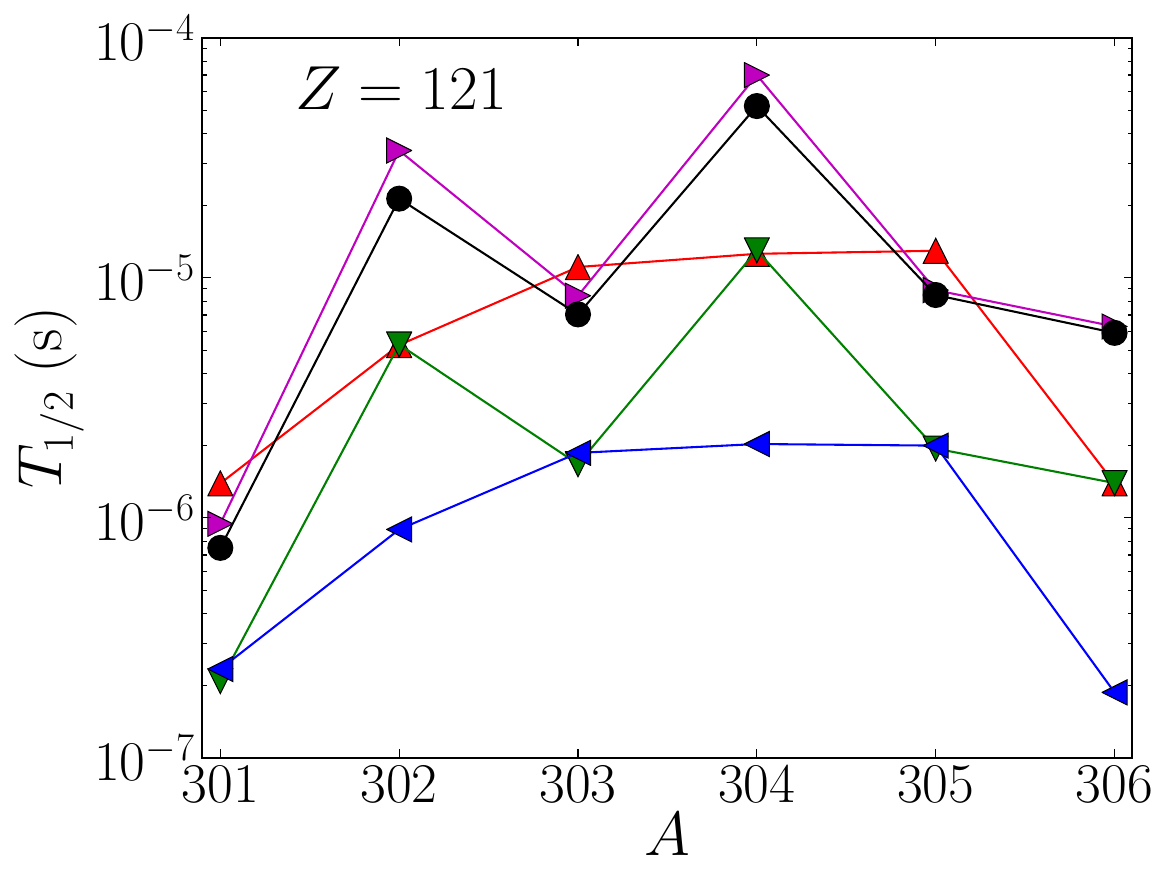}
\includegraphics[width=0.8\columnwidth]{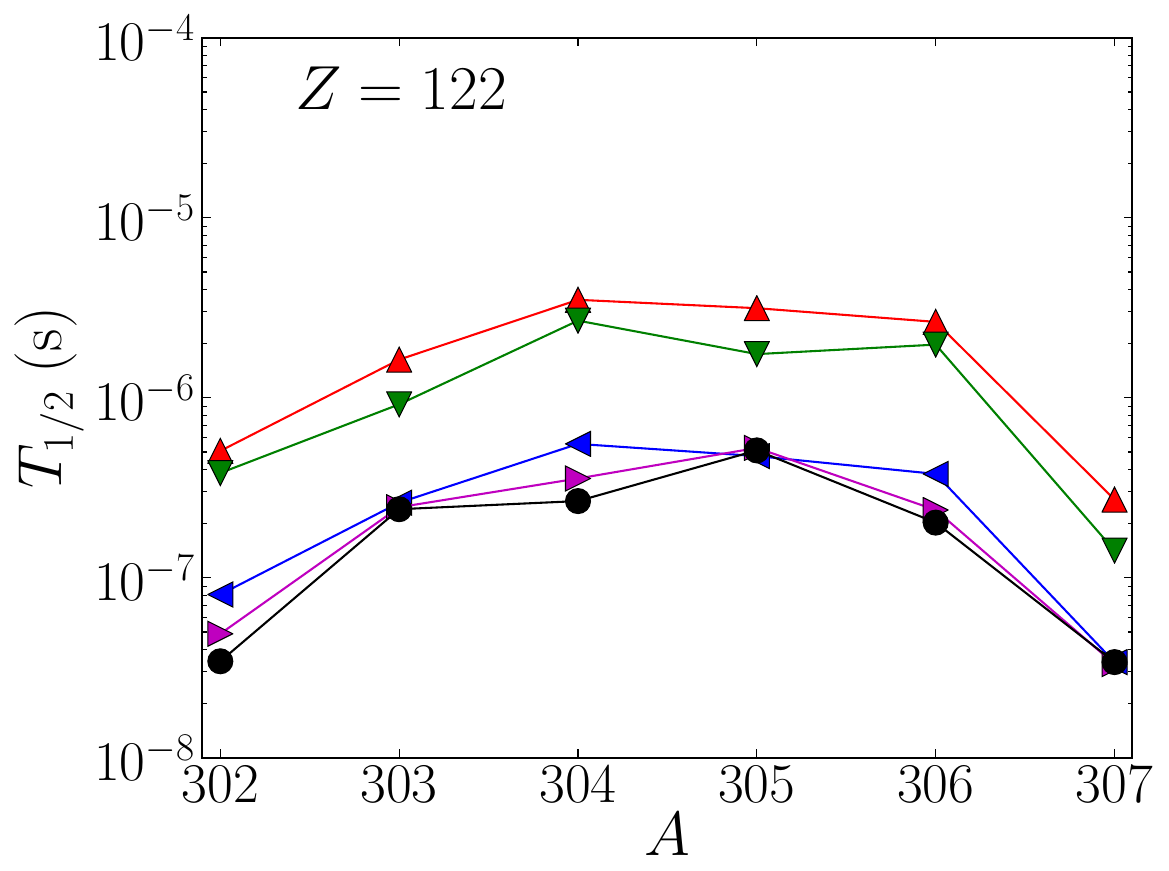}
\caption{Half-lives of superheavy elements listed in Table~\ref{tb:pre}.
Horizontal axis denotes the atomic mass number and the vertical axis represents the half-life
in units of seconds.}
\label{fig:superheavy}
\end{figure*}

The obtained $Q_\alpha$ values for heavy nuclei of $Z=117$--$122$ are listed in 
Table~\ref{tb:pre} together with their $\alpha$ decay half-lives predicted by the
SW, WS, Skyrme EDF potential models, and the VS formula. 
Here, VS and VS0 denote the VS formula with and without the isospin term, respectively.
The nuclei listed in Table~\ref{tb:pre} are along the valley of small $Q_\alpha$ values.
Because of the absence of the detailed information on their structure and quantum numbers, 
we simply assume $\ell = 0$.
Graphs shown in Fig.~\ref{fig:superheavy} visualize the half-lives listed in Table~\ref{tb:pre}.
For a given value of $Z$, the $\alpha$-decay lifetime actually depends on the $Q_\alpha$ value, 
and a longer lifetime is associated with a smaller $Q_\alpha$ value.
Comparing the results of the VS and VS0 formulas, we can see that the inclusion of the isospin term
increases lifetimes a little bit, but does not make significant difference.

\begin{table*}[t]
\caption{Theoretical predictions on $\alpha$ decay lifetimes of superheavy elements.  
The $Q_\alpha$ values are calculated with the WS4 mass table~\cite{WLWM14}. 
The modified and unmodified Viola-Seaborg formula are represented by VS and VS0, respectively.}
\begin{tabular}{c|cccccc}
\hline\hline
~Nuclei $(Z,A)$~  &  ~$Q_\alpha$ (MeV)~   &   $T_{1/2}^{\text{SW}}$ (s)  
        & ~$T_{1/2}^{\text{WS}}$ (s)~
        & ~$T_{1/2}^{\text{EDF}}$ (s)~
        & ~$T_{1/2}^{\text{VS}}$ (s)~  &  ~$T_{1/2}^{\text{VS0}}$ (s)~ \\
\hline
(122, 307) &  14.360  & $2.721\times 10^{-7}$    
           & $1.417\times 10^{-7}$   
           & $3.401\times 10^{-8}$   
           & $3.315\times 10^{-8}$   
           & $3.402\times 10^{-8}$  \\ 
(122, 306) &  13.775  & $2.641 \times 10^{-6}$  
           & $1.975\times 10^{-6}$ 
           & $3.777\times 10^{-7}$   
           & $2.380\times 10^{-7}$
           & $2.026\times 10^{-7}$ \\
(122, 305) &  13.734  & $3.147\times 10^{-6}$  
           &  $1.749\times 10^{-6}$
           & $4.746\times 10^{-7}$   
           & $5.266\times 10^{-7}$
           & $5.103\times 10^{-7}$ \\
(122, 304) &  13.710  & $3.503\times 10^{-6}$  
           &  $2.684\times 10^{-6}$ 
           & $5.544\times 10^{-7}$   
           & $3.563\times 10^{-7}$
           & $2.669\times 10^{-7}$ \\
(122, 303) &  13.904  & $1.630\times 10^{-6}$  
           &  $9.198\times 10^{-7}$ 
           & $2.614\times 10^{-7}$   
           & $2.468\times 10^{-7}$
           & $2.405\times 10^{-7}$ \\
(122, 302) &  14.208  & $5.069\times 10^{-7}$  
           &  $3.820\times 10^{-7}$ 
           & $8.078\times 10^{-8}$   
           & $4.887\times 10^{-8}$
           & $3.438\times 10^{-8}$ \\
\hline
(121, 306) &  13.783  & $1.392\times 10^{-6}$  
           & $1.396\times 10^{-6}$   
           & $1.873\times 10^{-7}$   
           & $6.268\times 10^{-6}$
           & $5.896\times 10^{-6}$ \\
(121, 305) &  13.242  & $1.296\times 10^{-5}$  
           & $1.943\times 10^{-6}$  
           & $1.999\times 10^{-6}$   
           & $8.881\times 10^{-6}$
           & $8.478\times 10^{-6}$ \\
(121, 304) &  13.251  & $1.259\times 10^{-5}$  
           &  $1.302\times 10^{-5}$ 
           & $2.030\times 10^{-6}$   
           & $6.994\times 10^{-5}$
           & $5.196\times 10^{-5}$ \\
(121, 303) &  13.283  & $1.109\times 10^{-5}$  
           & $1.673\times 10^{-6}$ 
           & $1.864\times 10^{-6}$   
           & $8.416\times 10^{-6}$
           & $7.039\times 10^{-6}$ \\
(121, 302) &  13.464  & $5.247\times 10^{-6}$  
           & $5.273\times 10^{-6}$   
           & $8.943\times 10^{-7}$   
           & $3.391\times 10^{-5}$
           & $2.137\times 10^{-5}$ \\
(121, 301) &  13.795  & $1.391\times 10^{-6}$  
           &  $2.086\times 10^{-7}$ 
           & $2.344\times 10^{-7}$   
           & $9.437\times 10^{-7}$
           & $7.494\times 10^{-7}$ \\
\hline
(120, 304) &  12.736  & $6.297\times 10^{-5}$  
           &  $4.862\times 10^{-5}$ 
           &  $1.041\times 10^{-5}$   
           & $7.245\times 10^{-6}$
           & $7.286\times 10^{-6}$ \\
(120, 303) &  12.782  & $5.151\times 10^{-5}$  
           &  $2.901\times 10^{-5}$ 
           &  $8.859\times 10^{-6}$   
           & $1.602\times 10^{-5}$
           & $1.509\times 10^{-5}$ \\
(120, 302) &  12.862  & $3.656\times 10^{-5}$ 
           &  $2.857\times 10^{-5}$ 
           &  $6.506\times 10^{-6}$   
           & $4.532\times 10^{-6}$
           & $4.074\times 10^{-6}$ \\
(120, 301) &  13.036  & $1.721\times 10^{-5}$  
           &  $9.805\times 10^{-6}$ 
           &  $3.117\times 10^{-6}$   
           & $4.612\times 10^{-6}$
           & $4.434\times 10^{-6}$ \\
(120, 300) &  13.290  & $5.916\times 10^{-6}$  
           &  $4.575\times 10^{-6}$ 
           & $1.075\times 10^{-6}$   
           & $7.214\times 10^{-7}$
           & $6.024\times 10^{-7}$ \\
(120, 299) &  13.230  & $7.675\times 10^{-6}$  
           &  $4.437\times 10^{-6}$   
           & $1.475\times 10^{-6}$   
           & $1.836\times 10^{-6}$
           & $1.784\times 10^{-6}$ \\
\hline
(119, 298) &   12.684 & $4.371\times 10^{-5}$  
           & $4.629\times 10^{-5}$  
           & $9.081\times 10^{-6}$   
           & $2.667\times 10^{-4}$
           & $1.950\times 10^{-4}$ \\
(119, 297) &   12.394 & $1.663\times 10^{-4}$  
           & $2.624\times 10^{-5}$  
           & $3.829\times 10^{-5}$   
           & $1.981\times 10^{-4}$
           & $1.486\times 10^{-4}$\\
(119, 296)  &   12.444 & $1.331\times 10^{-4}$  
           & $1.419\times 10^{-4}$  
           & $3.180\times 10^{-5}$   
           & $9.659\times 10^{-4}$
           & $5.733\times 10^{-4}$\\
(119, 295)  &   12.727 & $3.716\times 10^{-5}$  
           & $5.852\times 10^{-6}$ 
           & $8.815\times 10^{-6}$   
           & $4.307\times 10^{-5}$
           & $2.988\times 10^{-5}$\\
(119, 294)  &   12.695 & $4.332\times 10^{-5}$  
           & $4.460\times 10^{-5}$  
           & $1.084\times 10^{-5}$   
           & $3.657\times 10^{-4}$
           & $1.857\times 10^{-4}$\\
(119, 293) &   12.683 & $4.620\times 10^{-5}$  
           & $7.336\times 10^{-6}$ 
           & $1.208\times 10^{-5}$   
           & $6.122\times 10^{-5}$
           & $3.680\times 10^{-5}$\\
\hline
(118, 298) &  12.153  & $2.621\times 10^{-4}$  
           &  $2.108\times 10^{-4}$ 
           & $5.623\times 10^{-5}$   
           & $4.151\times 10^{-5}$
           & $4.030\times 10^{-5}$ \\
(118, 297) &  12.074  & $3.867\times 10^{-4}$  
           & $2.296\times 10^{-4}$
           & $1.967\times 10^{-4}$   
           & $1.893\times 10^{-4}$
           & $1.754\times 10^{-4}$ \\
(118, 296) &  11.722  & $2.232\times 10^{-3}$  
           & $1.894\times 10^{-3}$ 
           & $5.727\times 10^{-4}$   
           & $4.395\times 10^{-4}$
           & $3.546\times 10^{-4}$ \\
(118, 295) &  11.872  & $1.062\times 10^{-3}$  
           & $6.571\times 10^{-4}$ 
           & $2.772\times 10^{-4}$   
           & $5.688\times 10^{-4}$
           & $5.141\times 10^{-4}$\\
(118, 294) &  12.167  & $2.553\times 10^{-4}$  
           & $2.138\times 10^{-4}$ 
           & $6.608\times 10^{-5}$   
           & $4.991\times 10^{-5}$
           & $3.770\times 10^{-5}$ \\
(118, 293) &  12.210  & $2.103\times 10^{-4}$  
           & $1.307\times 10^{-4}$ 
           & $5.665\times 10^{-5}$   
           & $9.354\times 10^{-5}$
           & $8.666\times 10^{-5}$  \\
\hline
(117, 298) &  11.490  & $3.580\times 10^{-3}$  
           & $4.226\times 10^{-3}$   
           & $8.133\times 10^{-4}$   
           & $1.405\times 10^{-2}$
           & $1.693\times 10^{-2}$\\
(117, 297) &  11.589  & $2.162\times 10^{-3}$  
           & $3.478\times 10^{-4}$ 
           & $5.065\times 10^{-4}$   
           & $2.652\times 10^{-3}$
           & $2.762\times 10^{-3}$ \\
(117, 296) &  11.473  & $3.972\times 10^{-3}$  
           & $4.647\times 10^{-3}$  
           & $9.994\times 10^{-4}$   
           & $1.868\times 10^{-2}$
           & $1.840\times 10^{-2}$ \\
(117, 295) &  11.266  & $1.197\times 10^{-2}$  
           & $1.960\times 10^{-3}$ 
           & $3.285\times 10^{-3}$   
           & $1.911\times 10^{-2}$
           & $1.610\times 10^{-2}$ \\
(117, 294) &  11.346  & $7.897\times 10^{-3}$  
           &   $9.230\times 10^{-3}$   
           & $2.232\times 10^{-3}$   
           & $4.413\times 10^{-2}$
           & $3.526\times 10^{-2}$ \\
(117, 293) &  11.591  & ~$2.228\times 10^{-3}$~ 
           & $3.643\times 10^{-4}$  
           & $6.314\times 10^{-4}$   
           & $3.518\times 10^{-3}$
           &  $2.741\times 10^{-3}$  \\
\hline\hline
\end{tabular}\label{tb:pre}
\end{table*}

Among the nuclei in Table~\ref{tb:pre}, the lifetime of the (117, 294) nucleus was reported 
very recently~\cite{KYDA14}.
The reported experimental value of its lifetime is $54^{+94}_{-20}$~ms, which is about 20 times 
larger than our prediction of the Skyrme EDF model that gives $2.446$~ms.
We found that this discrepancy may be related to the difference of the $Q_\alpha$ value between 
the theoretical prediction and the measured value.
The WS4 model predicts $Q_\alpha = 11.346$~MeV~\cite{WLWM14}, but the measured value is
$11.20$~MeV~\cite{KYDA14}.
The difference is only about 1.3~\%, but as can be seen in the Geiger-Nutall law or the VS formula
of Eq.~(\ref{eq:VS}), the lifetime is very sensitive to the value of $Q_\alpha$ and one percent difference 
in $Q_\alpha$ could result in a factor of 10 difference in the lifetime.
This shows the sensitivity of the $\alpha$ decay lifetime to the nuclear structure and the important role 
carried by $Q_\alpha$ in determination of nuclear half-lives.

In fact, if we use the measured $Q_\alpha$ value in our calculation, the obtained lifetimes are in good
agreement with the measured lifetime as shown in Table~\ref{tb:117chain}, which also summarizes
the half-lives of the nuclei in the decay chain of the \nuclide[294]{117} nucleus.
In most cases the models we use in this work reproduce the experimental data as good as
in Table~\ref{tb:heavy}.
However, we note that the theoretical predictions overestimate the lifetime of the (113, 286) nucleus
by one or two orders of magnitude, which is similar to the observation mentioned in Ref.~\cite{QR14}.
More rigorous and complex analysis would be required to understand this discrepancy.
At the bottom of Table~\ref{tb:117chain}, therefore, we provide two sets of rms deviation values. 
The upper and lower rows represent the rms deviation values with and without the (113, 286) nucleus,
respectively.
Advantage of including the isospin-dependent term is evident when we compare the results of the VS 
and VS0 formulas except the isotope of (113, 286).

\begin{table*}
\caption{Half-lives of nuclides in the decay chain of the nucleus \nuclide[294]{117}. 
The experimental data are from Ref.~\cite{KYDA14}.}
\label{tb:117chain}
\begin{tabular}{ccccccccc}
\hline
\hline
$(Z,\,A)$  &  $Q_\alpha$  (MeV)
       & $T_{1/2}^\text{Exp}$ 
       & $T_{1/2}^{\text{SW}}$ 
       &  $T_{1/2}^{\text{WS}}$
       & $T_{1/2}^{\text{EDF}}$ 
       & $T_{1/2}^{\text{VS}}$ 
       & $T_{1/2}^{\text{VS0}}$ \\
\hline
(117,294) & ~$11.20 \, \pm 0.04 $~ 
          & ${51}^{+94}_{-20}$ ms 
          & ${17}^{+4}_{-3}$ ms 
          & ${34}^{+9}_{-7}$ ms 
          & ${22}^{+6}_{-4}$ ms 
          & ${96}^{+23}_{-18}$ ms 
          & ${75}^{+18}_{-14}$ ms \\ 
(115,290)  & $10.45 \,\pm 0.04$ 
          & ${1.3}^{+2.3}_{-0.5}$ s 
          & ${0.29}^{+0.08}_{-0.06}$ s 
          & ${2.0}^{+0.56}_{-0.44}$ s 
          & ${2.3}^{+0.64}_{-0.50}$ s 
          & ${1.40}^{+0.37}_{-0.29}$ s 
          & ${1.28}^{+0.33}_{-0.26}$ s  \\ 
(113,286)  & $9.4 \,\pm 0.3 $ 
          & ${2.9}^{+5.3}_{-1.1}$ s 
          & ${53}^{+398}_{-46}$ s 
          & ${71}^{+552}_{-62}$ s 
          & ${24}^{+191}_{-21}$ s 
          & ${208}^{+1452}_{-179}$ s 
          & ${209}^{+1390}_{-179}$ s  \\ 
(111,282) & $9.18 \,\pm 0.03$ 
          & ~${3.1}^{+5.7}_{-1.2}$ min~ 
          & ${0.81}^{+0.19}_{-0.16}$ min 
          & ${1.91}^{+0.46}_{-0.37}$ min  
          & ${1.96}^{+0.48}_{-0.38}$ min 
          & ${2.88}^{+0.66}_{-0.54}$ min 
          & ${3.60}^{+0.81}_{-0.66}$ min \\ 
(109,278)  & $ 9.59\,\pm 0.03$ 
          & ${3.6}^{+6.5}_{-1.4}$ s 
          & ${0.61}^{+0.13}_{-0.11}$ s 
          & ${4.70}^{+1.03}_{-0.84}$ s 
          & ${1.44}^{+0.32}_{-0.26}$ s
          & ${2.13}^{+0.45}_{-0.37}$ s 
          & ${3.63}^{+0.75}_{-0.62}$  \\ 
(107,274)  & $8.97 \,\pm 0.03$ 
          & ${30}^{+54}_{-12}$ s 
          & ${8.0}^{+1.9}_{-1.5}$ s 
          & ${18.8}^{+4.5}_{-3.6}$s  
          & ${22.9}^{+5.6}_{-4.5}$ s 
          & ${23.6}^{+5.5}_{-4.4}$ s 
          & ${48.0}^{+10.8}_{-8.8}$ s  \\ 
(105,270)  & $8.02\,\pm 0.03$ 
          & ${1.0}^{+1.9}_{-0.4}$ h 
          & ${0.57}^{+0.16}_{-0.12}$ h 
          & ${0.82}^{+0.23}_{-0.18}$ h  
          & ${0.39}^{+0.11}_{-0.09}$ h 
          & ${1.27}^{+0.35}_{-0.27}$ h 
          & ${2.91}^{+0.78}_{-0.61}$ h  \\ 
\hline
$\sigma$ &  &  & 0.769 & 0.592 & 0.486 & 0.773 & 0.790 \\
         &  &  & 0.625 & 0.185 & 0.340  & 0.173 & 0.241 \\
\hline
\hline
\end{tabular}
\end{table*}


\section{Conclusion}
\label{sec:summary}

The phenomenological potential for the $\alpha$ particle inside a nucleus and the WKB approximation
are the two key concepts to investigate $\alpha$ decay half-lives of nuclei in the cluster model.
In the present work, we propose to modify the nuclear potential of the $\alpha$ particle by
explicitly including the isospin-dependent terms containing $I=(N-Z)/A$ and we calculated 
the $\alpha$ decay half-lives of nuclei with the value of $I$ as large as $0.2$.
We also suggest a new effective potential of the $\alpha$ particle based on the Skyrme energy
density functional, which contains the isospin asymmetry contribution in a more natural way. 
Finally, we modified the empirical VS formula by including the $I$ and $I^2$ terms.

Although the $\alpha$ decay half-lives are mostly determined by the value of $Q_\alpha$, we found
that the isospin effects may improve the results to some extent as shown by our results. 
Together with the results of Ref.~\cite{DZS11}, which shows the importance of nuclear symmetry
energy in $Q_\alpha$ values, our findings indicate the important role of nuclear isospin asymmetry
effects in neutron-rich nuclei.

The potential model based on the Skyrme EDF suggests a form of the interaction between the
$\alpha$ particle and nucleon in the lowest order. 
The parameters of this approach are then obtained by fitting the $\alpha$-decay half-lives. 
In addition, the density profile of the core nucleus was found by the Thomas-Fermi 
approximation. 
The proposed EDF approach for $\alpha$ decay was found to explain successfully the decay events 
of heavy nuclei even better than the square well potential and Wood-Saxon potential approaches,
which may be ascribed to the realistic density profile of the core nucleus based on a microscopic 
approach.
This indicates that the isospin asymmetry may alter the penetration length of the potential barrier as well.

In the present work, we first parameterize the nuclear potential of the $\alpha$ particle and fit 
the parameters to the data. 
Therefore, in this process, we cannot take into account the specific properties of each nucleus.
As a result, the effects which come from, for example, shell structure, deformation, preformation factor
of $\alpha$ particle could not be properly taken into account.
Therefore, improving the present model calculations along this direction and inclusion of isospin
asymmetry effects in microscopic models would be desired to better understand nuclear $\alpha$ decay
of neutron-rich nuclei.


\acknowledgments

We are grateful to N. Itagaki for fruitful discussions.
This work was supported in part by the Basic Science Research Program through the National Research 
Foundation (NRF) of Korea funded by the Ministry of Education under Grant 
Nos.~NRF-2013R1A1A2A10007294, NRF-2014R1A1A2054096, and NRF-2015R1D1A1A01059603. 
The work of Y.L. was supported by the Rare Isotope Science Project funded by the Ministry of Science, 
ICT and Future Planning (MSIP) of the Korean Government and the National Research 
Foundation (NRF) of Korea under Grant No. 2013M7A1A1075766.

\appendix*
\section{}

In microscopic approaches, the $\alpha$ particle bound state with a nucleus can be studied 
by solving the Hartree-Fock equation.
As given in Eq.~(\ref{eq:skyint}), we start with the potential in the form of
\begin{widetext}
\begin{eqnarray}
v_{N\alpha}^{} (\bm{k},\bm{k}') &=& s_0^{} \left( 1 + v_0^{} P_\sigma \right) 
\delta ( \bm{r}_{N\alpha}^{} )
+ \frac{s_1^{}}{2} \left( 1+ v_1^{} P_\sigma \right) \left[ \delta(\bm{r}_{N\alpha}^{}) \bm{k}^2 
+ \bm{k}'^2 \delta(\bm{r}_{N\alpha}) \right]  
+ s_2^{} \, \bm{k}' \cdot \delta(\bm{r}_{N\alpha}^{}) \bm{k}  
\nonumber \\ && \mbox{}
+ i \, W_0^{\alpha} \bm{k}'  \cdot \left( \bm{\sigma} \times \bm{k} \right) \delta(\bm{r}_{N\alpha}^{}) 
+ \frac{s_3^{}}{6} \left( 1 + v_3^{} P_\sigma \right) \rho_N^{\epsilon}\, \delta(\bm{r}_{N\alpha}^{}) \,.
\label{eq:skyint2}
\end{eqnarray}
When kinetic energy is included, the above interaction leads to the Hamiltonian for $\alpha$ particle
as
\begin{eqnarray}
\mathcal{H}_\alpha &=& \frac{\hbar^2}{2m_\alpha^{}} \tau_\alpha^{}
+ s_0^{}  \left(1 + \frac{v_0^{}}{2} \right) \rho_N^{} \rho_\alpha^{} 
+ \frac{1}{4} \left( s_1^{} + s_2^{} \right) \left(\tau_\alpha^{} \rho_N^{} 
+ \tau_N^{} \rho_\alpha^{} \right) 
+ \frac{1}{4} \left( 3 s_1^{} - s_2^{} \right) 
\left(\bm{\nabla} \rho_N^{} \cdot \bm{\nabla} \rho_\alpha^{} \right) 
\nonumber \\ && \mbox{}
+ \frac{1}{4}\, s_3^{} \, \rho_N^{\epsilon}\, \rho_\alpha^{} \left( \rho_N^2 
+ 2 \rho_n^{} \rho_p^{} \right) 
+ \frac{1}{2} \, W_0^\alpha \left( \bm{\nabla} \rho_N^{} \cdot  \bm{J}_\alpha 
+ \bm{\nabla} \rho_\alpha^{} \cdot \bm{J}_N \right) ,
\end{eqnarray}
where $\tau$ and $\bm{J}$ are expressed as 
\begin{equation}
\tau_A^{}(\bm{r})  = \sum_{i} \left\vert \bm{\nabla} \varphi_i \right\vert^2\,,
\quad
\bm{J}_A(\bm{r})  = \sum_{i} \varphi_i^{\dagger} \left(-i \bm{\nabla} \times \bm{\sigma} \right)
\varphi_i^{} \,,
\end{equation}
for the nucleon ($A=N$) and the $\alpha$ particle ($A=\alpha$). 
The single particle wave function $\varphi(\bm{r})$ of the $\alpha$ particle can be obtained by solving 
the Hartree-Fock equation. 
In the spherically symmetric case, the wave function can be written as
\begin{equation}
\varphi_i^{}(\bm{r}) = \frac{R_{n \ell j}(r)}{r}
\braket{ \ell \, m_\ell^{} \, s \, \sigma \mid j\, m} Y^\ell_m(\hat{\bm{r}}),
\end{equation} 
and the Schr\"{o}dinger equation becomes
\begin{eqnarray}
\left[-\frac{d}{dr}\frac{\hbar^2}{2m_\alpha^*}\frac{d}{dr}
+ \frac{\hbar^2}{2m_\alpha^*}\frac{\ell(\ell + 1)}{r^2} + V_{N}(r)  \right] R_{n \ell j}^{}(r) 
= e_{n \ell j}^{} R_{n \ell j}^{}(r) \,,
\end{eqnarray}
where the potential for the $\alpha$ particle reads
\begin{eqnarray}
V_{N}(\bm{r}) & = & s_0^{} \left(1 +\frac{1}{2} v_0^{} \right) \rho_N^{} 
+ \frac{1}{4} \left( s_1^{} + s_2^{} \right) \left(\tau_p^{} + \tau_n^{} \right) 
- \frac{1}{4}(3 s_1^{} - s_2^{}) \rho_N^{\prime\prime}
- \left( \frac{5}{4} s_1^{} - \frac{3}{4} s_2^{} \right) \frac{\rho_N^\prime}{r} 
\nonumber \\ && \mbox{} 
+ \frac{1}{4} s_3^{} \rho_N^\epsilon \left(\rho_N^2 + 2 \rho_n^{} \rho_p^{} \right)
- \frac{W_0^\alpha}{2} \left( J_N^\prime + \frac{2}{r} J_N \right) 
+ \frac{1}{2} W_0^\alpha \, \frac{\rho_N^\prime}{r}
\left[j(j+1) -\ell(\ell+1) - \textstyle\frac{3}{4}\right]\,,
\end{eqnarray}
\end{widetext}
where $\rho_N^{} = \rho_p^{} + \rho_n^{}$ with $\rho_N^\prime = d \rho_N^{} /dr$ 
and $\rho_N^{\prime\prime} = d^2 \rho_N^{} /dr^2$.
The effective mass $m_\alpha^*$ is defined by
\begin{equation}
\frac{\hbar^2}{2m_\alpha^*} 
= \frac{\hbar^2}{2m_\alpha^{}} + \frac{1}{4} \left( s_1^{} + s_2^{} \right) \rho_N^{} \,.
\end{equation}
Since the total spin of the $\alpha$ particle is zero, i.e., $\braket{\sigma_\alpha} = 0$,
the spin-orbit coupling between the $\alpha$ particle and the daughter nucleus
may be neglected.
This process leads to the form of the effective potential of the $\alpha$ particle as
\begin{eqnarray}
V_{N} &=& \alpha \rho_N^{} + \beta \left( \rho_n^{5/3} + \rho_p^{5/3} \right)
+ \gamma\, \rho_N^\epsilon (\rho_N^2 + 2 \rho_n^{} \rho_p^{})
\nonumber \\ && \mbox{}
+ \delta \frac{\rho_N^\prime}{r} + \eta \rho_N^{\prime\prime}\,,
\end{eqnarray}
where we have used that $\tau_{p,n}^{} \simeq  \frac{3}{5}(3\pi^2)^{2/3} \rho_{p,n}^{5/3}$ 
within the Thomas-Fermi approximation.



\begin{thebibliography}{10}

\bibitem{Mang64}
H.~J. Mang,
\newblock Alpha decay,
\newblock Ann. Rev. Nucl. Sci. \textbf{14}, 1 (1964).

\bibitem{GSI}
Facility for Antiproton and Ion Research in Europe GmbH,
\url{http://www.fair-center.eu}

\bibitem{FRIB}
FRIB Users Organization,
\url{http://fribusers.org/frib/science.html}

\bibitem{HM00}
S.~Hofmann and G.~Munzenberg,
\newblock The discovery of the heaviest elements,
\newblock Rev. Mod. Phys. \textbf{72}, 733 (2000).

\bibitem{LCK08}
B.-A. Li, L.-W. Chen, and C.~M. Ko,
\newblock Recent progress and new challenges in isospin physics with heavy-ion
  reactions,
\newblock Phys. Rep. \textbf{464}, 113 (2008).

\bibitem{BMP91}
B.~Buck, A.~C. Merchant, and S.~M. Perez,
\newblock Ground state to ground state alpha decays of heavy even-even nuclei,
\newblock J. Phys. G \textbf{17}, 1223 (1991).

\bibitem{BMP92}
B.~Buck, A.~C. Merchant, and S.~M. Perez,
\newblock Favoured alpha decays of odd-mass nuclei,
\newblock J. Phys. G \textbf{18}, 143 (1992).


\bibitem{BMP92b}
B.~Buck, A.~C. Merchant, and S.~M. Perez,
\newblock $\alpha$ decay calculations with a realistic potential,
\newblock Phys. Rev. C \textbf{45}, 2247 (1992).

\bibitem{Royer00}
G.~Royer,
\newblock Alpha emission and spontaneous fission through quasi-molecular shapes,
\newblock J. Phys. G \textbf{26}, 1149 (2000).

\bibitem{DZWZL10}
J.~Dong, H.~Zhang, Y.~Wang, W.~Zuo, and J.~Li,
\newblock Alpha-decay for heavy nuclei in the ground and isomeric states,
\newblock Nucl. Phys. A \textbf{832}, 198 (2010).

\bibitem{RSB05}
P.~Roy~Chowdhury, C.~Samanta, and D.~N. Basu,
\newblock $\alpha$ decay half-lives of new superheavy elements,
\newblock Phys. Rev. C \textbf{73}, 014612 (2006).

\bibitem{SCB07}
C.~Samanta, P.~R. Chowdhury, and D.~N. Basu,
\newblock Predictions of alpha decay half lives of heavy and superheavy
  elements,
\newblock Nucl. Phys. A \textbf{789}, 142 (2007).

\bibitem{RSB08}
P.~Roy~Chowdhury, C.~Samanta, and D.~N. Basu,
\newblock Search for long lived heaviest nuclei beyond the valley of stability,
\newblock Phys. Rev. C \textbf{77}, 044603 (2008).

\bibitem{DI05}
V.~Yu. Denisov and H. Ikezoe,
\newblock $\alpha$-nucleus potential for $\alpha$-decay and sub-barrier fusion,
\newblock Phys. Rev. C \textbf{72}, 064613 (2005).

\bibitem{LZZS10}
L.-L. Li, S.-G. Zhou, E.-G. Zhao, and W. Scheid,
\newblock A new barrier penetration formula and its application to alpha-decay half-lives,
\newblock Int. J. Mod. Phys. E \textbf{19}, 359 (2010).

\bibitem{DZS11}
J.~Dong, W.~Zuo, and W.~Schied,
\newblock Correlation between $\alpha$-decay energies of superheavy nuclei
  involving the effects of symmetry energy,
\newblock Phys. Rev. Lett. \textbf{107}, 012501 (2011).

\bibitem{DZG12}
J.~Dong, W.~Zuo, and J.~Gu,
\newblock Origin of symmetry energy in finite nuclei and density dependence of
  nuclear matter symmetry energy from measured $\alpha$-decay energies,
\newblock Phys. Rev. C \textbf{87}, 014303 (2013).

\bibitem{RHL11}
C.-Y. Ryu, C.~H. Hyun, and C.-H. Lee,
\newblock Hyperons and nuclear symmetry energy in neutron star matter,
\newblock Phys. Rev. C \textbf{84}, 035809 (2011).

\bibitem{GCR11}
S.~Gandolfi, J.~Carlson, and S.~Reddy,
\newblock Maximum mass and radius of neutron stars, and the nuclear symmetry  energy,
\newblock Phys. Rev. C \textbf{85}, 032801(R) (2012).

\bibitem{TSCD12}
M.~B. Tsang \textit{et~al.\/},
\newblock Constraints on the symmetry energy and neutron skins from experiments and theory,
\newblock Phys. Rev. C \textbf{86}, 015803 (2012).

\bibitem{LL12}
J.~M. Lattimer and Y.~Lim,
\newblock Constraining the symmetry parameters of the nuclear interaction,
\newblock Astrophys. J. \textbf{771}, 51 (2013).

\bibitem{DSS13}
C.~Drischler, V.~Som{\`a}, and A.~Schwenk,
\newblock Microscopic calculations and energy expansions for neutron-rich  matter,
\newblock Phys. Rev. C \textbf{89}, 025806 (2014).

\bibitem{GK87b}
S.~A. Gurvitz and G.~Kalbermann,
\newblock The decay width and the shift of a quasistationary state,
\newblock Phys. Rev. Lett. \textbf{59}, 262 (1987).

\bibitem{ASN97}
S.~\r{A}berg, P.~B. Semmes, and W.~Nazarewicz,
\newblock Spherical proton emitters,
\newblock Phys. Rev. C \textbf{56}, 1762 (1997),
\newblock Phys. Rev. \textbf{58}, 3011(E) (1998).

\bibitem{WS54}
R.~D. Woods and D.~S. Saxon,
\newblock Diffuse surface optical model for nucleon-nuclei scattering,
\newblock Phys. Rev. \textbf{95}, 577 (1954).

\bibitem{VTMLC91}
R.~L. Varner, W.~J. Thompson, T.~L. McAbee, E.~J. Ludwig, and T.~B. Clegg,
\newblock A global nucleon optical model potential,
\newblock Phys. Rep. \textbf{201}, 57 (1991).

\bibitem{CBHMS98}
E.~Chabanat, P.~Bonche, P.~Haensel, J.~Meyer, and R.~Schaeffer,
\newblock A Skyrme parametrization from subnuclear to neutron star densities
  Part II. Nuclei far from stabilities,
\newblock Nucl. Phys. A \textbf{635}, 231 (1998).

\bibitem{Langer37}
R.~E. Langer,
\newblock On the connection formulas and the solutions of the wave equation,
\newblock Phys. Rev. \textbf{51}, 669 (1937).

\bibitem{VB72}
D.~Vautherin and D.~M. Brink,
\newblock Hartree-Fock calculations with Skyrme's interaction. I. Spherical
  nuclei,
\newblock Phys. Rev. C \textbf{5}, 626 (1972).

\bibitem{AWWK12}
G.~Audi, M.~Wang, A.~H. Wapstra, F.~G. Kondev, M.~MacCormick, X.~Xu, and
  B.~Pfeiffer,
\newblock The \textsc{Ame2012} atomic mass evaluation (I). Evaluation of input
  data, adjustment procedures,
\newblock Chin. Phys. C \textbf{36}, 1287 (2012).

\bibitem{WAWK12}
M.~Wang, G.~Audi, A.~H. Wapstra, F.~G. Kondev, M.~MacCormick, X.~Xu, and
  B.~Pfeiffer,
\newblock The \textsc{Ame2012} atomic mass evaluation (II). Tables, graphs and
  references,
\newblock Chin. Phys. C \textbf{36}, 1603 (2012).

\bibitem{GN11}
H.~Geiger and J.~M. Nuttall,
\newblock The ranges of the $\alpha$ particles from various substances and a
  relation between range and period of transformation,
\newblock Phil. Mag. Ser. 6 \textbf{22}, 613 (1911).

\bibitem{VS66b}
V.~E. Viola, Jr. and G.~T. Seaborg,
\newblock Nuclear systematics of the heavy elements - II. Lifetimes for alpha,
  beta and spontaneous fission decay,
\newblock J. Inorg. Nucl. Chem. \textbf{28}, 741 (1966).

\bibitem{RXW04}
Z.~Ren, C.~Xu, and Z.~Wang,
\newblock New perspective on complex cluster radioactivity of heavy nuclei,
\newblock Phys. Rev. C \textbf{70}, 034304 (2004).

\bibitem{RZ08}
G.~Royer and H.~F. Zhang,
\newblock Recent $\alpha$ decay half-lives and analytic expression predictions
  including superheavy nuclei,
\newblock Phys. Rev. C \textbf{77}, 037602 (2008).

\bibitem{NRDX08}
D.~Ni, Z.~Ren, T.~Dong, and C.~Xu,
\newblock Unified formula of half-lives for $\alpha$ decay and cluster
  radioactivity,
\newblock Phys. Rev. C \textbf{78}, 044310 (2008).

\bibitem{SPC89}
A.~Sobiczewski, Z.~Patyk, and S.~\'{C}wiok,
\newblock Deformed superheavy nuclei,
\newblock Phys. Lett. B \textbf{224}, 1 (1989).

\bibitem{SSP15}
K.~P. Santhosh, I.~Sukumaran, and B.~Priyanka,
\newblock Theoretical studies on the alphs decay of \nuclide[178-220]{Pb}
  isotopes,
\newblock Nucl. Phys. A \textbf{935}, 28 (2015).

\bibitem{OULA06}
\mbox{Yu}.~\mbox{Ts}. Oganessian \textit{et~al.\/},
\newblock Synthesis of the isotopes of elements 118 and 116 in the
  \nuclide[249]{Cf} and $\nuclide[249]{Cm} + \nuclide[48]{Ca}$ fusion
  reactions,
\newblock Phys. Rev. C \textbf{74}, 044602 (2006).

\bibitem{OULA04b}
\mbox{Yu}.~\mbox{Ts}. Oganessian \textit{et~al.\/},
\newblock Measurements of cross sections and decay properties of the isotopes
  of elements 112, 114, and 116 produced in the fusion reactions
  \nuclide[233,238]{U}, \nuclide[242]{Pu}, and $\nuclide[248]{Cm} +
  \nuclide[48]{Ca}$,
\newblock Phys. Rev. C \textbf{70}, 064609 (2004),
\newblock \textbf{71}, 029902(E) (2005).

\bibitem{OULA04-OUDL05}
\mbox{Yu}.~\mbox{Ts}. Oganessian \textit{et~al.\/},
\newblock Experiments on the synthesis of element 115 in the reaction
  $\nuclide[243]{Am}(\nuclide[48]{Ca}, xn) \nuclide[291-x]{115}$,
\newblock Phys. Rev. C \textbf{69}, 021601(R) (2004),
\newblock \textbf{69}, 029902(E) (2004);
\mbox{Yu}.~\mbox{Ts}. Oganessian \textit{et~al.\/},
\newblock Synthesis of elements 115 and 113 in the reaction $\nuclide[243]{Am}
  + \nuclide[48]{Ca}$,
\newblock Phys. Rev. C \textbf{72}, 034611 (2005).

\bibitem{OULA07}
\mbox{Yu}.~\mbox{Ts}. Oganessian \textit{et~al.\/},
\newblock Synthesis of the isotope \nuclide[282]{113} in the $\nuclide[237]{Np}
  + \nuclide[48]{Ca}$ fusion reaction,
\newblock Phys. Rev. C \textbf{76}, 011601(R) (2007).

\bibitem{DZGWP10}
J.~Dong, W.~Zuo, J.~Gu, Y.~Wang, and B.~Peng,
\newblock $\alpha$-decay half-lives and $Q_\alpha$ values of superheavy nuclei,
\newblock Phys. Rev. C \textbf{81}, 064309 (2010).

\bibitem{WLWM14}
N.~Wang, M.~Liu, X.~Wu, and J.~Meng,
\newblock Surface diffuseness correction in global mass formula,
\newblock Phys. Lett. B \textbf{734}, 215 (2014).

\bibitem{MNMS93}
P.~M{\"o}ller, J.~R. Nix, W.~D. Myers, and W.~J. Swiatecki,
\newblock Nuclear ground-state masses and deformations,
\newblock Atom. Data Nucl. Data Tabl. \textbf{59}, 185 (1995).

\bibitem{DZ94}
J.~Duflo and A.~P. Zuker,
\newblock Microscopic mass formulae,
\newblock Phys. Rev. C \textbf{52}, 23 (1995).

\bibitem{GCP10}
S.~Goriely, N.~Chamel, and K.~M. Pearson,
\newblock Further exploration of Skyrme-Hartree-Fock-Bogoliubov mass formulas.
  XII: Stiffness and stability of neutron-star matter,
\newblock Phys. Rev. C \textbf{82}, 035804 (2010).

\bibitem{MRDT06}
E.~L. Medeiros, M.~M.~N. Rodrigues, S.~B. Duarte, and O.~A.~P. Tavares,
\newblock Systematics of alpha-decay half-life: New evaluations for
  alpha-emitter nuclides,
\newblock J. Phys. G \textbf{32}, B23 (2006).

\bibitem{KYDA14}
J.~Khuyagbaatar \textit{et~al.\/},
\newblock $\nuclide[48]{Ca} + \nuclide[249]{Bk}$ fusion reaction leading to
  element $Z=117$: Long-lived $\alpha$-decaying \nuclide[270]{Db} and discovery
  of \nuclide[266]{Lr},
\newblock Phys. Rev. Lett. \textbf{112}, 172501 (2014).

\bibitem{QR14}
Y.~Qian and Z.~Ren,
\newblock Half-lives of $\alpha$ decay from natural nuclides and from
  superheavy elements,
\newblock Phys. Lett. B \textbf{738}, 87 (2014).

\end{thebibliography}
\end{document}